\documentclass{emulateapj}

\slugcomment{accepted by the Astrophysical Journal}

\begin{document}

\title{The Rest-Frame FUV Morphologies of Star-Forming Galaxies at $z \sim 1.5$ and $z \sim 4$}

\author{Jennifer M. Lotz \altaffilmark{1}, Piero Madau \altaffilmark{1},
Mauro Giavalisco \altaffilmark{2}, Joel Primack \altaffilmark{3}, 
and Henry C.Ferguson \altaffilmark{2}}

\altaffiltext{1}{Department of Astronomy and Astrophysics, 
University of California, Santa Cruz, CA 95064; jlotz@ucolick.org, 
pmadau@ucolick.org}
\altaffiltext{2}{Space Telescope Science Institute,
3700 San Martin Dr., Baltimore, MD 21218; mauro@stsci.edu, 
ferguson@stsci.edu}
\altaffiltext{3}{Department of Physics, 
University of California, Santa Cruz, CA 95064; joel@physics.ucsc.edu}

\begin{abstract}
We apply a new approach to quantifying galaxy morphology and identifying
galaxy mergers to the rest-frame far-ultraviolet images
of 82 $z \sim 4$ Lyman break galaxies (LBGs) and 55 $1.2 < z < 1.8$ emission-line galaxies
in the GOODS and Ultra Deep Fields.  We compare the
distributions of the Gini coefficient ($G$), 
second-order moment of the brightest 20\% of galaxy light ($M_{20}$), 
and concentration ($C$) for high-redshift and low-redshift galaxies 
with average signal to noise per pixel $> 2.5$ and Petrosian
radii $> $0.3\arcsec.  
Ten of the 82 LBGs have $M_{20}  \ge -1.1 $  and possess bright double or multiple nuclei,
implying a major-merger fraction of star-forming galaxies $\sim$ 10-25\% at 
$M_{FUV} < -20$, depending
on our incompleteness corrections.  Galaxies with bulge-like morphologies
($G \ge 0.55$, $M_{20} < -1.6$) make up $\sim$ 30\% of the $z \sim 4$ LBG sample,
while the remaining $\sim$ 50\% have $G$ and $M_{20}$ values higher than expected for
smooth bulges and disks and may be star-forming disks, minor-mergers or post-mergers. 
The star-forming $z \sim 1.5$ galaxy sample has a morphological distribution
which is similar to the UDF $z \sim 4$ LBGs, with an identical fraction of 
major-merger candidates but fewer spheroids.  The observed morphological distributions
are roughly consistent with current hierarchical model predictions for the
major-merger rates and minor-merger induced starbursts at $z \sim 1.5$ and $\sim 4$. 
We also examine the rest-frame $FUV-NUV$ and $FUV-B$ colors 
as a function of morphology and find no strong correlations at
either epoch.
\end{abstract}

\keywords{galaxies:evolution -- galaxies:high-redshift -- galaxies:interacting -- 
galaxies:structure -- ultraviolet:galaxies}

\section{INTRODUCTION}
Galaxy interactions and mergers are expected to
play a dominant role in the formation of stars and 
evolution of galaxies, particularly in the early universe.
The present hierarchical models of galaxy assembly and star formation
fail to predict enough red spheroidal galaxies -- the assumed 
end products of major galaxy mergers (e.g. Toomre 1977; Mihos \& Hernquist 1996) 
-- at both early and late times (e.g. Cimatti et al. 2004). 
Observations of on-going mergers at high redshift 
and the earliest spheroidal systems are crucial to 
constraining the role of mergers in the evolution of galaxies
and the build-up of stellar mass.
One of the most accessible ways of identifying galaxy mergers
and their end-products is by their morphologies.

Morphological studies of very distant objects have become possible
in the last decade with the high-resolution capabilities of
the {\it Hubble Space Telescope}.  Galaxies at redshifts
greater than one appear to be systematically 
more disturbed than local galaxies (Abraham et al. 1996), 
implying strong evolution in the number density of merging
and interacting systems (Abraham \& van den Bergh 2002).
The first {\it HST} images of galaxies above $z \sim 3$ identified
via the Lyman break technique (Steidel et al. 1995; Madau et al. 1996) 
showed a ranges of morphologies, from compact
proto-bulges to disturbed major-merger candidates (Giavalisco, 
Steidel, \& Macchetto 1996; Lowenthal et al. 1997; Dickinson 1999; Giavalisco 2002). 
In recent years, high redshift objects have been identified by other techniques
and subsequently studied with {\it HST}.   These include the
extremely red objects (EROs) and sub-mm sources at $z \sim 1-3$.
EROs possess a range of morphologies from spheroidal to irregular
(Moustakas et al. 2004),  with an increasing fraction with
multiple nuclei at $z > 2$ (Daddi et al. 2003).
Optical {\it HST} images of a small sample of sub-mm sources show that
70\% possess multiple components (Chapman et al. 2003).

The connection between the morphologies of very early objects
and those observed at later times is not straightforward.
In the current picture of galaxy evolution, 
a galaxy's morphology is not constant through cosmic time, nor
does it evolve along a single linear path. Star-forming disks are 
generally grown 
through the accretion of cold intergalactic gas, while
bulges and spheroids are typically produced in mergers and interactions.
For example, in one simulation of the
evolution of a present-day elliptical, its progenitors  
start as disks grown at redshifts  $>$ 4 (Steinmetz \& Navarro 2002).  
These primordial disk galaxies merge to
form a bulge at $z \sim 3$.  The merger remnant then dynamically relaxes,
accretes more gas and reforms a disk between $1 < z < 2$.  At 
$z < 1$, the galaxy undergos yet more interactions and 
minor-mergers until its gas supply 
is exhausted and it resembles a red elliptical at $z \sim 0$.  

Despite the complex evolutionary path a single object may take,
the evolution of the morphological distribution of galaxies and
the number of merging systems at $z > 1$ is an important tracer
of the formation of the Hubble sequence.  While disks are
quite fragile and can be destroyed and then regrown, numerical simulations
suggest that, once created, bulges are not transient and 
only grow in mass as a galaxy evolves (Mihos \& Hernquist 1996, 
Steinmetz \& Navarro 2002).  At $z \geq 3$, 
the majority of galaxies are predicted to be disk-dominated
(Somerville, Primack, \& Faber 2001, hereafter SPF01). 
However, many of these proto-disks may not resemble local spirals
because the disks might not be stable (Mao, Mo, \& White 1998) 
and because a substantial fraction are expected to be undergoing 
minor mergers (SPF01).  Also, by $z \sim 3$,
approximately 20\% of objects may have already transformed into bulge-dominated
systems (SPF01). The fraction of dark matter halos
undergoing a major merger is expected to increase as a power-law 
proportional to $(1+z)^3$ out to at least $z \sim 2$.  Above $z \sim 2$,
theoretical predictions imply that the merger rate flattens such
that $\sim 20-30\%$ of massive dark matter halos at $z > 2$ should merge within a
Gyr (Gottl\"ober, Klypin, \& Kravtsov 2001).  
High resolution hydrodynamical simulations indicate
that the associated gas-rich galaxy mergers create bulge-dominated galaxies
(e.g. Mihos \& Hernquist 1996).

Some of these mergers should evolve into the $z \leq 1$ red spheroidal 
population, particularly if they live in dense environments.
The spatial clustering of high redshift objects has been used
to argue that bright Lyman break galaxies are the likely progenitors of
highly-clustered present-day ellipticals (Giavalisco et al. 1994; Giavalisco 2002), although
fainter LBGs are much less clustered (Giavalisco \& Dickinson 2001) and
alternative models are possible (e.g. Wechsler et al. 2001).  Other $z \geq 2$
galaxy populations are also highly clustered (e.g. Daddi et al. 2001).  
Yet it remains ambiguous to what extent the observed high redshift galaxies track the
progenitors of local spheroids, since their 
stellar masses, metallicities, and star-formation rates must 
evolve substantially at later times (Dickinson et al. 2003;
Ferguson et al. 2002; Lowenthal et al. 1997).

Despite the great advances in identifying high-redshift galaxies
and obtaining high-resolution images, the distribution of intrinsic
galaxy morphologies at early times is still uncertain.
Complicating the issue is the wavelength dependence on 
observed morphology.  The highest spatial resolution {\it HST}
images are those taken in optical wavelengths with the
Advanced Camera for Surveys (ACS).  At $z > 1.2$, 
ACS can only sample rest-frame wavelengths less than the
4000\AA\ break, where dust and young star-forming regions
dominate the morphologies of local galaxies (Giavalisco et al. 1996; 
Hibbard and Vacca 1997; Goldader et al. 2002).  Therefore the
morphological distributions of high-redshift objects classified by
their rest-frame ultraviolet morphologies should not
be compared to the rest-frame optical morphological distributions
of nearby objects. 
 
A second problem is the intrinsic difficulty of measuring 
the morphologies of faint and compact objects like Lyman break
galaxies.  Lyman break galaxies at $z \sim 4$ have typical 
rest-frame far-ultraviolet (FUV) half-light radii between 
0.1 and 1.0 \arcsec\ (Ferguson et al. 2004; Bouwens et al. 2004), 
therefore many LBGs with sizes below at few tenths of an
arcsecond are not spatially resolved enough
to measure their morphologies even in ACS images. 
The compact sizes may bias standard morphological
analyses that are commonly applied to lower redshift objects,
such as concentration measures and bulge-disk decompositions.
Also, images of faint high-redshift objects are
likely to miss light in the outer, lower-surface brightness
regions of the objects.  This missed light will bias the
observed sizes low, and also affect the measured concentrations
and surface-bright profiles.  Finally, morphology measures like
rotational asymmetry and clumpiness which involve subtracting images
and/or a correction for the background noise signal are less robust
at low signal-to-noise levels (Lotz, Primack, \& Madau 2004).

A third possible problem is the assumption that disturbed
morphologies at high redshift directly correlate with mergers and 
interactions.  Because of the roughly 500 million year timescales 
associated with merger activity, we do not truly know if a 
disturbed object has undergone a recent merger from a snapshot of 
its morphology alone.  The observed morphology depends on the
observational effects (rest-frame wavelength, viewing angle, dust
extinction) and the physical conditions of the merger/interaction
(mass ratio, orbital parameters, gas fractions, bulge fractions, 
internal kinematics). Spectroscopy can sometimes confirm the presence of two
orbiting galaxies (e.g. the Superantenna, Mihos \& Bothun 1998) or tidal in-fall 
(e.g. IRAS 1520+3342, Arriaba \& Colina 2002).  Recent observations
of morphological irregular, elongated systems at $z \sim 2$ have
failed to show the significant rotation that is expected at early merger stages
(Erb et al. 2004), and recent studies of kinematic close pairs at $z \sim 1$
indicate a lower merger rate than that derived using morphological 
asymmetry (Lin et al. 2004). However, 
the detection of  kinematic merger signatures will depend on
the merger stage as well as viewing angle and the slit-position.  
ULIRGs in the final merger stage can exhibit complex velocity fields 
without obvious rotational centers (eg. IRAS 23128-5919, Mihos \& Bothun 1998).   
While this paper assumes that morphologically identified double-nucleated galaxies 
are merging systems, theoretical and empirical calibrations of quantitative 
morphologies and kinematic signatures 
as merger indicators are important tasks for future work.

In this paper, we apply a new approach to quantifying galaxy
morphology and identifying mergers at high redshift.
First presented in Lotz, Primack and Madau (2004; hereafter LPM04),   
this technique measures the relative distribution of
galaxy pixel flux values or Gini coefficient (Abraham et al. 2003) and the 
second-order moment of the brightest 20\% of the galaxy's
pixels ($M_{20}$).  These two quantities are extremely robust to
decreasing signal-to-noise and change less than 10\%
for average signal-to-noise per pixel $> 2$ at spatial resolutions better
than 500 pc per resolution element (LPM04).  At rest-frame optical
wavelengths, we found that local galaxy mergers separate cleanly from the local galaxy
Hubble types in Gini coefficient ($G$) and $M_{20}$.  

Here we present the non-parametric rest-frame 
FUV morphologies for 82 $z \sim 4$ Lyman break galaxies
observed in the GOODS and UDF survey and compare the
morphology distributions to a spectroscopically selected
sample of 55 $z \sim 1.5$ galaxies in the GOODS fields.
In \S 2, we discuss the selection of both galaxy samples and note
that our analysis is limited to FUV bright star-forming galaxies
at both epochs.  In \S 3 we describe our non-parametric measures,
and how these are applied to the GOODS and UDF images.  The size and surface-brightness
selection effects are examined in \S 4 with both a comparison of 
UDF and GOODS-South morphologies and a series of simulations. 
The observed morphological distributions of objects with sufficient
sizes and signal-to-noise are presented in \S5.  We also compare the
galaxy morphologies to their colors for a sub-set of objects in the
GOODS-South and UDF.  Finally, in \S6 we discuss the morphological
distributions in the context of hierarchical model predictions for
star-forming galaxies at $z \ge 1.5$.

\section{GALAXY SAMPLES}
Images with high spatial resolutions (better than 1 kpc per resolution element) 
are needed to reliably measure galaxy morphologies. Assuming the
concordance cosmological model ($H_o = 70 $ km s$^{-1}$ Mpc$^{-1}$, 
$\Omega_m = 0.3, \Omega_\Lambda = 0.7$), galaxies at redshifts greater 
than $z \sim 1$ have angular scales $\sim$ 5 kpc per arcsec.  
The point spread function (PSF) of the Advanced Camera for Surveys (ACS) on
the Hubble Space Telescope ($HST$) has a 
FWHM $= 0.12~\arcsec$, or $\ge 800$ pc at $z \ge 1$.  
Two recent ACS programs have provided an archive of deep, high-resolution,
multi-wavelength images that are invaluable for the study of very distant galaxies:
Great Observatories Origins Deep Survey (GOODS; Giavalisco et al. 2004) 
and the Ultra Deep Field (PI: S. Beckwith).
The GOODS ACS images cover two 160 \sq\arcmin\ fields (the Hubble Deep Field
North and the Chandra Deep Field South) in four band-passes (F435W,
F606W, F775W, and F850W), with a 10 $\sigma$ detection limit of $\sim$ 26.7 in 
F850W.   The UDF covers a much smaller area ($\sim$ 11.3 \sq\arcmin) in the Chandra
Deep Field South, but is observed in the same ACS band-passes and
has a deeper limiting magnitude ($\sim$ 28.4 in F850W). 
An additional advantage of these two
surveys is that the overlap between the extremely deep UDF and somewhat shallower 
GOODS fields allows us to empirically determine the limiting signal-to-noise 
for our morphology measures.  

We use these two sets of ACS images 
to measure the morphologies of Lyman break galaxies (LBGs) at $z \sim 4$ 
and a lower-redshift comparison sample at $z \sim 1.5$.   At $z \sim 4$, even
the longest wavelength $z$-band (with a central wavelength $\lambda_c = 8500$\AA)
ACS images probe light emitted at rest-frame far-ultraviolet (FUV) wavelengths.  
Longer wavelength NICMOS images do exist for the UDF and the HDFN, but these 
have significantly worse resolution (NIC3 PSF FWHM $\ge$ 0.22 - 0.35 \arcsec).   
We will restrict our analysis to the FUV
morphologies of the Lyman break galaxies and the comparison sample
of lower-redshift galaxies at  $1.2 < z < 1.8$.

\subsection{$z \sim 4$ Lyman-break Galaxies}
The $z \sim 4$ LBGs were selected from the GOODS photometric catalogs 
using the ``dropout'' technique (eg. Madau et al. 1996, Steidel et al. 1996),  via 
the $B_{435}$-dropout color-criterion of Giavalisco et al. (2004). 
This color-cut is sensitive to star-forming galaxies with low to moderate extinctions and
redshifts $3.4 \le z \le 4.5$ with an expected median redshift $z \sim 3.9$.
The deepest GOODS images are in the $V$ (F606W) and
$i$ (F775W) bands, which are sufficiently redward of the Lyman break at $z \sim 4$
to detect and quantify the LBG morphologies. 
We have chosen to examine the LBG morphologies
in the summed $V$ and $i$ ACS images, in order to boost the 
signal-to-noise per pixel for each galaxy and obtain the largest possible sample
of LBGs with measurable morphologies.  The effective
wavelength of these summed images is 6272 \AA, or rest-frame 1425 \AA\ at
$z = 3.9$.  We have compared the sizes and quantitative morphologies of 
high signal-to-noise LBGs measured in the $V+i$ images to those measured in summed
$i+z$ images, and have found no evidence for patchy Lyman-alpha forest
absorption in the $V$ band images which might have affected the apparent morphologies. 
As high spatial resolution is not necessary to obtain reliable photometry, we also utilize the 
{\it European Southern Observatory} ($ESO$) VLT/ISAAC ground-based near-infrared photometric survey 
of the GOODS-S (CDFS) fields (Vandame et al. 2004) to compare the rest-frame FUV morphologies
to rest-frame $FUV-B$ colors (\S 5.2).

Each LBG candidate brighter than 26.6 mag in the
summed $V+i$ images was visually inspected, and objects which were spurious detections 
(i.e. saturated star diffraction spikes, cosmic-rays) 
were removed from the sample.  A fraction of the LBG candidates 
are compact and red enough to be late-type M stars. However, 
as discussed below, all objects with $r_p < 0.3\arcsec$ 
were removed for our morphological analysis and we expect no 
stellar contamination for our subsample of LBGs. 
The GOODS LBG catalog has $\sim$ 660 B-dropouts detected above 10 $\sigma$ in the
summed $V+i$ images ($\sim 26.6$ mag).  The smaller but deeper UDF LBG 
catalog has 77 B-dropouts detected above 10 $\sigma$ in the summed $V+i$ image
($\sim 27.9$ mag). 

\subsection{ $z \sim 1.5$ galaxies}
At $z \sim 1.2-1.8$, the bluest GOODS ACS $B$ (F435W) images
sample the rest-frame FUV at $\sim$ 1980 - 1550 \AA, slightly
redder than $V+i$ images for the $B$-dropouts.  Identifying objects 
in this redshift range is notoriously difficult. 
Spectroscopic redshifts are difficult to obtain because few 
emission lines or spectral breaks are redshifted into the observed 
optical wavelengths, and optical color selection fails because 
the Lyman break is bluer than the optical band-passes. 
Photometric redshifts suffer from similar problems and we found that
the incidence of catastrophic failure ($|(z_{spec} -z_{phot})|/ (1 + z_{spec}) \ge 0.2$) 
in this redshift range was 25\%.
Nevertheless, spectroscopic confirmation for some GOODS $z \sim 1.5$
galaxies has been obtained by several surveys.  The public Treasury Keck
Redshift Survey with DEIMOS on the Keck II telescope has 
obtained redshifts for 1440 objects out of a $R_{AB}=24.4 $ magnitude-limited 
sample of 2018 galaxies in the GOODS-North field (Wirth et al. 2004).  The 
TKRS catalog has ``good'' quality redshifts for 112 galaxies between $1.2 < z < 1.8$. 
The VIMOS VLT Deep Survey has publicly released redshifts for 1599 objects
with $I_{AB} \leq 24$ mag in the GOODS-South (CDFS) field obtained with
VIMOS at the $ESO$ Very Large Telescope (Le F\`evre et al. 2004).  
Out of the VIMOS VLT Deep Survey catalog objects, 
68 galaxies had good quality spectra with $1.2 < z < 1.8$.
Finally, the $ESO$/GOODS spectroscopic program with the FORS2 spectrograph
at the VLT has obtained spectroscopic redshifts for 243 objects with $z_{850} < 24.5$ mag
in the GOODS-South field (Vanzella et al. 2004), and provides an additional 53 objects
with high quality redshifts between $1.2 < z < 1.8$ to our sample. 
The morphologies for all 408 $1.2 < z < 1.8$ objects were measured in the
GOODS ACS $B$ images.  

We note that all of these spectroscopic surveys are biased towards star-forming
galaxies with strong emission lines.
Spectroscopic redshifts are much more difficult to obtain for 
red galaxies at these redshifts, therefore red or weak emission-line
galaxies are likely to be missing from this sample.

\section{MORPHOLOGICAL ANALYSIS}
\subsection{Pre-processing}
The GOODS and UDF ACS images are large mosaics containing thousands of
galaxies. In order to identify the galaxies of interest and mask out foreground/background
objects, we ran the galaxy detection and photometry software SExtractor
(Bertin \& Arnouts 1996).  Segmentation maps were created for each 
mosaic image ($V+i$ or $B$)  using SExtractor with  the 
detection threshold set to 0.6 $\sigma$ and a minimum detection area of 16 pixels. 
The resulting segmentation maps were used as masks for 
foreground/background objects. 

Some of the LBGs have very close neighbors. Bright
star-forming regions of local galaxies may appear as several distinct
objects when viewed in the FUV (e.g. Goldader et al. 2002).  To prevent the artificial
break-up of our LBGs, we compared the SExtractor segmentation 
map for each LBG with the $B-V$ and  $V-i$ colormaps.  
If the pixels of a nearby object meeting the LBG color criteria overlapped
the pixels assigned to the LBG, or if another LBG from the GOODS
B-dropout catalog was within 1.5~\arcsec, the segmentation map was edited to include 
that object's pixels with the original LBG.  This affects 4 objects in our
final UDF LBG sample and 7 objects in the final GOODS LBG sample.

Postage stamps for each selected galaxy were cut out from 
the GOODS $B$ or  $V+i$ images and the SExtractor segmentation maps.   
The pixels assigned to any foreground 
objects in the postage stamp were masked out.  
A local sky region was selected to exclude any pixels flagged by the
SExtractor segmentation map, and the sky value within that box was
subtracted from the postage stamp. 

\subsection{Morphology Measurements}
The elliptical Petrosian radius, Gini coefficient, $M_{20}$, concentration, 
and average signal-to-noise per pixel ($\langle S/N \rangle$) were measured from the
cleaned, sky-subtracted postage stamp of each galaxy.  

A well-defined map of the galaxy's pixels is needed to compute $G$ and $M_{20}$.  
We choose not to use the isophotal-based SExtractor segmentation 
map. Rather, we recomputed our own map based on the surface brightness of the galaxy
at the Petrosian radius ($r_p$), where $r_p$ is
measured in elliptical apertures (see LPM04 for a discussion). The elliptical
Petrosian radius was calculated assuming the SExtractor ELLIPTICITY and
THETA values, and using the SExtractor computed XCENTER and YCENTER as
a first guess.  Galaxy pixels with fluxes greater than the
surface brightness at this radius were assigned to the segmentation map of the galaxy.  
Occasionally this new segmentation map is not congruous, and the
code fails. This generally occurs when either the
background object masking is not sufficient or when the object has a low surface
brightness relative to the sky level (i.e. low $\langle S/N \rangle$). 
For the $z \sim 1.5$ and $z\sim 4$ samples, any code failures were masked by hand and 
re-analyzed.

A new estimate of the galaxy's center was then determined by minimizing the total
second-order moment of the pixels within this map:
\begin{equation}
M_{total} = \sum_i^n M_i = \sum_i^n f_i \cdot ((x_i - x_c)^2 + (y_i - y_c)^2)
\end{equation}
where $f_i$ is the flux in each galaxy pixel and $x_c$, $y_c$ is 
the galaxy center. We recalculated $r_p$ using the revised 
center, and a revised segmentation map was used to calculate the Gini coefficient
and $M_{20}$.  The Gini coefficient is defined as the 
distribution of galaxy pixel flux values: 
\begin{equation}
G = \frac{1}{\bar{|X|} n (n-1)} \sum^n_i (2i - n -1) |X_i|
\end{equation}
where $n$ is the total number of pixels in the galaxy segmentation map.
and $X_i$ are the rank-ordered pixel flux values (Glasser 1962).
The second order moment of the brightest 20\% of the galaxy pixels is
defined as 
\begin{eqnarray}
M_{20} \equiv {\rm log10}\left(\frac{\sum_i M_i}{M_{tot}}\right) & {\rm with } & \sum_i f_i <  0.2 f_{tot}
\end{eqnarray}
Concentration is calculated from the ratio of radii of
circular apertures containing 80\% and 20\% of the total flux, 
where the total flux is the flux within 1.5 circular Petrosian radii (Conselice 2003) 
about the galaxy center determined by Eqn. 1.  Previous definitions of
concentration have been calculated about the asymmetry center (LPM04) or SExtractor
center (Conselice 2003). We do not calculate
asymmetry here due to its instability at low $\langle S/N \rangle$ (LPM04), 
and instead use the more reliable second-order moment center.
A bad choice of the center can lead to large systematic errors in the
concentration for very compact objects.

The average signal-to-noise per pixel ($\langle S/N \rangle$) 
was also computed for the pixels in our segmentation map (as computed above), 
using the variance of sky pixel values from the previously selected sky 
region:
\begin{equation}
\langle S/N \rangle = \frac{1}{n} \sum_i^n \frac{f_i}{\sqrt{\sigma_{sky}^2 + f_i}}
\end{equation}
Because the GOODS ACS images are drizzled and remapped to a pixel scale 
less than the original scale, the noise in the sky is correlated.  
We corrected $\langle S/N \rangle$ for the effects of 
drizzling using the variance reduction factor $F_A$ given in Eqn. A20 of 
Casertano et al. (2000):
\begin{equation}
\sigma_{correct} = \sqrt{\frac{\sigma_{sky}^2}{F_A}}
\end{equation}  
$F_A$ depends on the ratio $s$ of the output pixel scale to the original pixel scale 
and the drizzled pixel fraction $p$.  The GOODS images have $s = 0.6$ and  $p = 0.7$,
giving $F_A$ (GOODS) = 0.375.  The UDF pixel fraction $p$ is 0, therefore
$F_A$ (UDF) = 1.0.

\section{SELECTION EFFECTS AND BIASES}
Lyman break galaxies are among the most distant galaxies known, and 
the images we observe (even with the $HST$ ACS) are often low signal-to-noise
and barely resolved.  Any analysis of the morphologies of such high-redshift
objects must also include a careful study of the limits of the morphological measures
used.  In LPM04, we tested the stability of $G$, $M_{20}$, and $C$ with decreasing
$\langle S/N \rangle$ and spatial resolution for local galaxies observed in
rest-frame $R$.  First, the images for set of 7 galaxies of 
different morphological types were held a fixed spatial resolution (better than
120 pc per resolution element) but added to increasingly higher Poison noise images
to measure the effect of noise on the morphological quantities. Then the images 
were smoothed to decreasing spatial resolutions but scaled such that 
$\langle S/N \rangle$ was constant to measure the effect of low spatial resolution on 
the morphological quantities.  These tests showed that for relatively nearby galaxies
observed in rest-frame optical wavelengths, $G$ and $M_{20}$ can be recovered
for observations with $\langle S/N \rangle > 2-3$ and spatial resolutions better
than 1 kpc.  

However, other effects may come into play when measuring the morphologies
of distant galaxies.  The local sample of LPM04 may not adequately reflect the range of
morphologies or physical scales of objects at $z \ge 1$.  High redshift galaxies
may be appear more irregular, due to both evolution and morphological K-corrections
(Abraham et al. 1996).  Also, high redshift galaxies are physically smaller 
than local galaxies (e.g. Ferguson et al 2004).  Finally, the morphologies of galaxies 
smaller than a few times the image's PSF cannot be measured, because most of the 
object's spatial information is lost. 

In this work, we use three approaches to 
to determine the morphological selection effects and biases of our distant galaxy samples: 
1) a comparison of the measured morphologies for large sample of galaxies observed
by both the GOODS survey and the deeper UDF survey; 2) 
a series of simulations of face-on smooth exponential disks and 
bulges with varying sizes and magnitude; and 3) simulations in which we 
artificially redshift our $z \sim 1.5$ sample to $z= 4$.  The first test will help us 
understand at what $\langle S/N \rangle$ level the measured morphologies of real 
(and morphologically complex) distant galaxies are reproducible.
The second test will help determine any offsets from 
the ``true'' morphologies as a function of size, surface
brightness, and light profile.  The last test will allow us to directly compare the
observed morphological distributions at $z \sim 1.5$ and $z\sim4$.

\begin{figure*}
\plotone{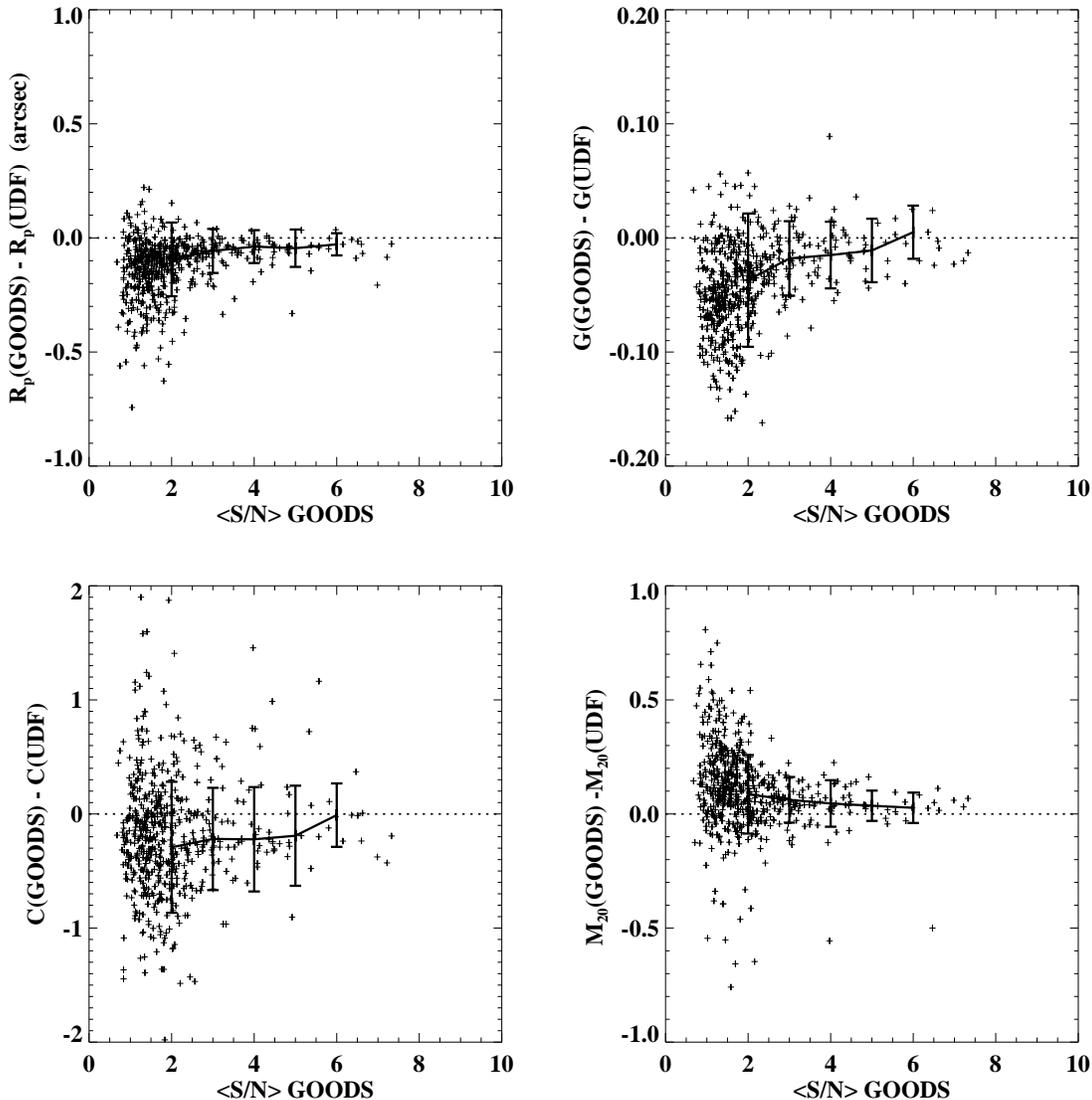}
\caption{GOODS - UDF morphologies for 571 galaxies in $V+i$ images}
\end{figure*}

\subsection{GOODS v. UDF morphologies}
One of the advantages of the GOODS data-set is that a sub-sample of the GOODS
CDFS field has been re-observed with $HST$ ACS in the same bandpasses to a substantially
greater depth for the Ultra Deep Field Survey.  
By comparing the size, magnitudes, and morphologies calculated from
the extremely high signal-to-noise images from the Ultra Deep Field survey
to those calculated for the exact same objects in the shallower GOODS images, 
we can learn a great deal about the reliability of the morphologies measured
for the entire GOODS sample. 

We have selected $\sim$ 750 objects from the UDF catalog with $z$ magnitudes brighter
than $z (F850W) = 28$ mag as our control sample.  
No color or redshift cut was applied to this sample, and
it includes galaxies as bright as $z (F850W) = 20$ mag.  
Both the UDF and overlapping GOODS summed $V+i$
images were pre-processed as described above without modification to the SExtractor
segmentation maps, and the morphologies and $\langle S/N \rangle$ for
all 750 objects were measured in both the GOODS and UDF images.  From this sample,
we selected galaxies with UDF $\langle S/N \rangle$ $\ge$ 3.0 and 
UDF elliptical Petrosian radii $r_p \ge 0.3$ \arcsec. 
We also removed objects for which the morphology code failed, i.e. objects for the
Gini segmentation map was not congruous.  The morphology
code failed for less than 5\% of galaxies brighter than $V+i$ = 26.5 in the GOODS-S images.

The differences between the GOODS and UDF
morphology and size measurements for the resulting 571 objects are plotted as a
function of the GOODS $\langle S/N \rangle$ in Figure 1. The median offset (GOODS-UDF) and
standard deviations for the 67 objects with $2.5 \le \langle S/N \rangle$(GOODS) $< 3.5$
is 
\begin{eqnarray}
\begin{array}{lll}
\Delta r_p & = -0.057 \arcsec   & \pm\ 0.097 \arcsec \\
\Delta G   & = -0.018           & \pm \ 0.033 \\
\Delta M_{20} & = + 0.062      & \pm \ 0.100 \\
\Delta C      & = -0.219          & \pm \ 0.448
\end{array} 
\end{eqnarray}
We repeated this analysis for the $B (F435W)$ GOODS and UDF images and
found very similar offsets and standard deviations at $2.5 \le \langle S/N \rangle$(GOODS) $< 3.5$.
At $\langle S/N \rangle$(GOODS) $< 2.5$, the differences between the measured
UDF and GOODS morphologies show large scatter.   Therefore, we will restrict
our analysis of galaxy morphologies to objects with $\langle S/N \rangle \ge 2.5$

At low $\langle S/N \rangle$, the Petrosian radii measured in the 
GOODS images are slightly smaller than those measured in the UDF images.  
This is likely to be the result of light in the outer low surface
brightness regions being lost in the sky noise.  This bias toward smaller sizes is probably
the reason for a small offset in $G$ at low $\langle S/N \rangle$. 
The lowest surface brightness pixels are not included in the segmentation map, 
and the Gini coefficient is artificially lowered.   The same bias may also be
the cause for the mean offset to higher $M_{20}$ values at low $\langle S/N \rangle$.  
$M_{20}$ is the ratio of the second order moment of light for the bright regions of the galaxy to the
total second order moment for the entire galaxy.  This total moment depends on the Petrosian
radius;  if the Petrosian radius is artificially lower, then the total moment decreases while
the moment of the brightest region remains unaffected.   
The concentrations measured in the GOODS images appear to be systematically lower than
those measured in the UDF images, even at relatively high $\langle S/N \rangle$. 
This offset ($-0.22$) is small relative to the dispersion (0.45) and the range
of possible concentrations. This is also the result of missed light in the
outer regions of the galaxy -- we measure systematically smaller
apertures containing 80\% of the total light in the GOODS images, even for
galaxies with $\langle S/N \rangle \sim 4$.

\subsection{Bulge and Disk Simulations}
To determine the robustness of
the measured morphologies at smaller sizes and fainter fluxes, 
we have simulated a series of face-on exponential 
disks and $r^{1/4}$ bulges of varying sizes and magnitudes.
The simulated galaxies are constructed using IRAF.ARTDATA.MKOBJECTS
task and are added to a blank region of the $V+i$ or $B$ images. 
We detect and measure the simulated objects in the same way as for our
real galaxy samples.  A noise-free image of de Vaucouleurs bulge 
with a half-light radius of 600 resolution elements has $G= 0.600$, $M_{20} =
-2.47$, and $C=4.34$, while a noise-free image of an exponential disk with the
same half-light radius has $G=0.473$, $M_{20} = -1.80$, and $C=2.71$
(horizontal dotted lines in Figure 2).

\begin{figure*}
\plotone{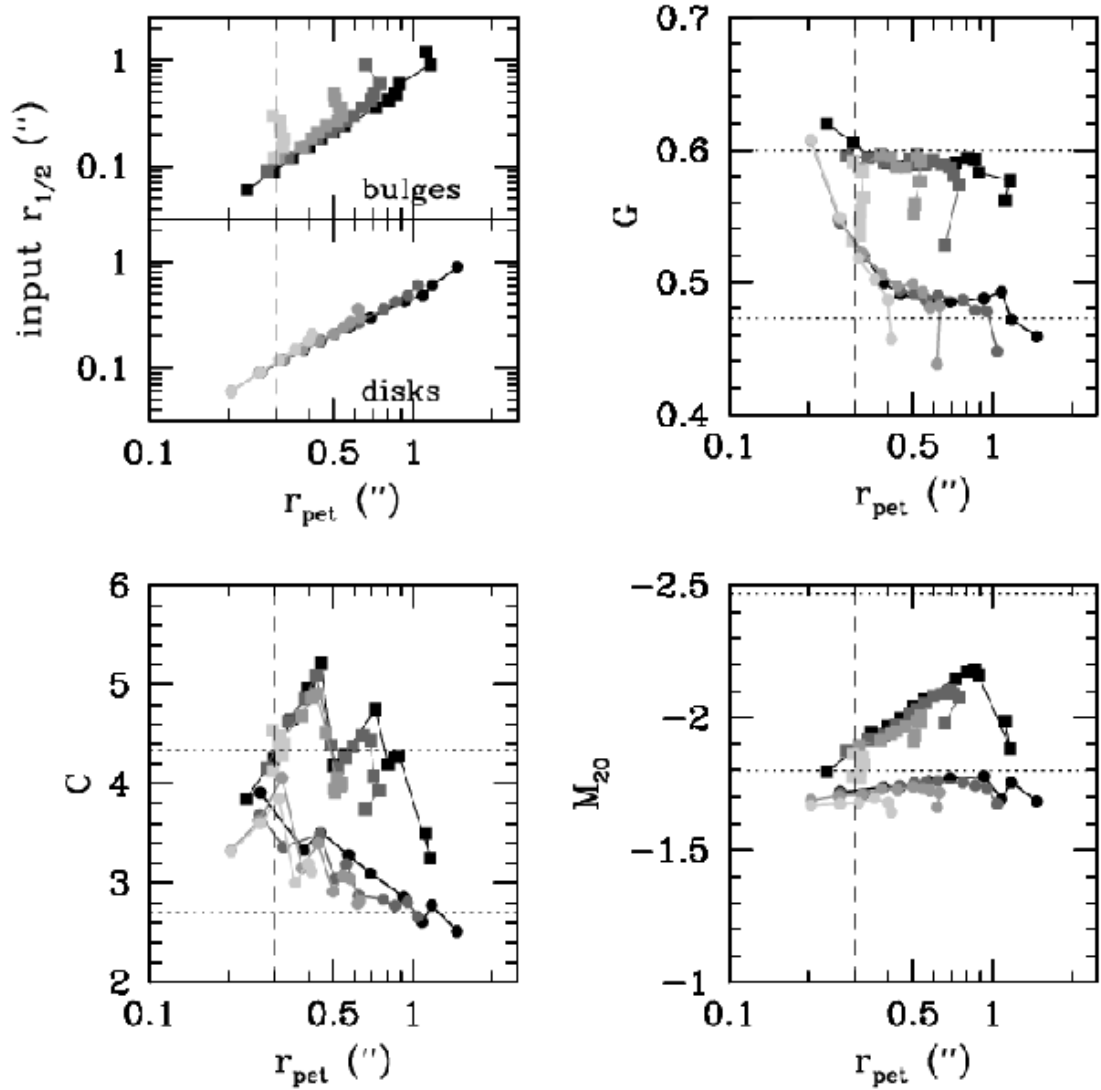}
\caption{ The measured morphologies as a function of measured
size ($r_p$) and magnitude for 
simulated face-on exponential disks (circles) and de Vaucouleurs bulges (boxes).
The input total magnitudes are in 1 magnitude steps, ranging from 25 (light grey)
to 22 (black). The horizontal dotted lines show the ``true'' morphological values.
The dashed line show the $r_p = 0.3$\arcsec cutoff.}.
\end{figure*}

We use these simulations to estimate any size/luminosity dependent 
biases in the morphology measurements. At $z \ge 3$, the ACS images 
have physical scales $\sim 600$ pc per PSF FWHM (0.12~\arcsec).  Therefore, we expect that intrinsically 
small and/or highly concentrated galaxies for which the central regions are not resolved 
will have systematic biases in their measured morphologies. 
We find that $C$, $M_{20}$, and to a lesser extent
$G$ are affected.  In Figure 2, we plot the measured morphologies and
sizes for both bulges and disks detected with $\langle S/N \rangle \ge 2.5$, 
with different grey-scales indicating different total input magnitudes.  
Our simulations show that disk-dominated galaxies 
with $r_p < 0.3$\arcsec\ are not sufficiently 
resolved to give accurate $G$. At small angular sizes, 
the measured $G$ approaches that expected for a point source ($G \sim 0.7$).   
We find that $M_{20}$ is biased low for the bulges with sizes less than 1~\arcsec; 
local well-resolved spheroids have $M_{20} \sim -2.5$ (LPM04), while our simulations produce
$M_{20} \sim -2.0$. Nevertheless, bulges and disks still have distinguishable $M_{20}$ values. 
At large sizes and faint magnitudes, we find that the measured radii are smaller than
their true radii as the flux from the outer low surface brightness regions is
lost.  This is a problem particularly for bulges. 
Because the lowest surface brightness pixels are lost, this effect 
can also artificially lower the measured $G$ by 0.05-0.06.

Concentration is unreliable at small sizes, because the inner aperture
containing 20\% of the total light not resolved.  The observed step-like
behavior of concentration for the simulated bulges at $r_p < 0.7$\arcsec\ 
is a result of interpolating the inner 20\% light radius ($r_{20}$) 
at less than 1 or 2 pixels.   At radii $<$ 1 pixel, the interpolated $r_{20}$ 
is likely to be an overestimate because of the steep inner surface 
brightness profile of  $r^{1/4}$ bulges, and results in an underestimate
of $C$.   The estimated $r_{20}$ approaches the true value at $r_{20}$ = 
1 pixel, and then is again overestimated at $r_{20} >$ 1 pixel.  
Once the measured $r_{20}$ is greater than 2 pixels, this effect is 
not as significant. 

In addition to selecting galaxies with $\langle S/N \rangle \ge 2.5$, we add the
additional selection criterion that the galaxy must have measured
$r_p> 0.3$ \arcsec, because smaller objects are likely to 
have artificially high $G$ values.  This will also effectively eliminate
any possible stellar contaminants in our LBG sample. 
The net result is that our sample will span a limited range in sizes,
as larger objects are likely to have too low $\langle S/N \rangle$ 
and smaller objects are too compact to be measured. 
In Figure 3, we show the measured $r_p$ and isophotal magnitudes for
the GOODS and UDF B-dropouts and the GOODS $z \sim 1.5$ sample as a function
of $\langle S/N \rangle$.  All panels in Figure 3 are plotted on the same
physical scale. The high $\langle S/N \rangle$ $z \sim 1.5$ 
galaxies are intrinsically larger 
and fainter than the GOODS B-dropouts,
but overlap in absolute magnitude and size with the UDF B-dropouts.
The redshift distribution of the selected $z \sim 1.5$ subsamples is shown in Fig. 4.

\begin{figure*}
\plotone{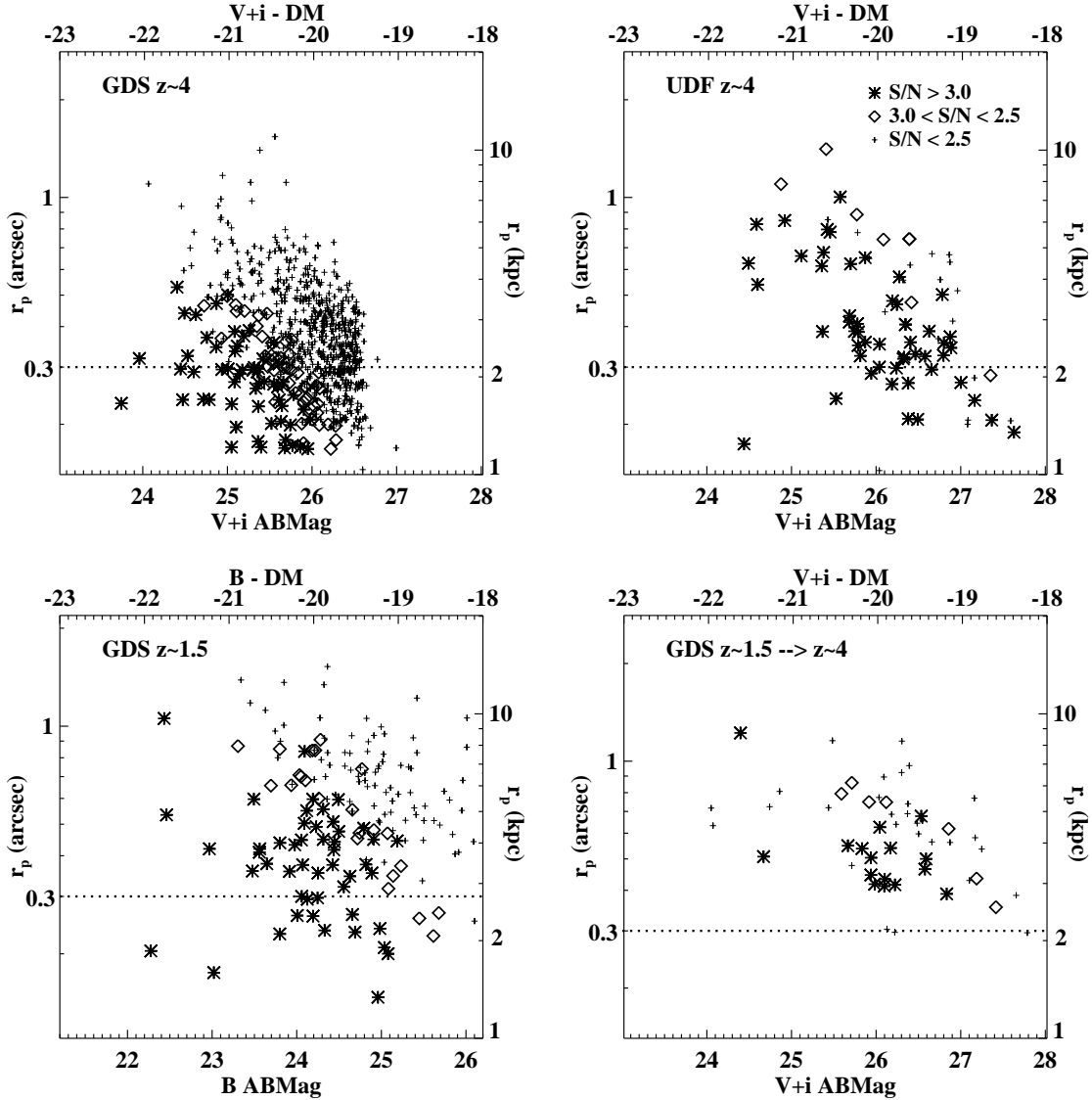}
\caption{ $r_p$ v. isophotal magnitude for GOODS and UDF B-dropouts (upper panels)
$z \sim 1.5$ sample and artificially redshifted $z \sim 1.5 \rightarrow 4$ sample (lower panels).
All four panels are set to the same physical scale in absolute magnitude and intrinsic size 
(kpc).}
\end{figure*}

\subsection{Artificial redshift simulations}
In Figure 3, we demonstrated that the UDF $z \sim 4$ and GOODS $z \sim 1.5$ samples
span a similar range of intrinsic sizes and rest-frame FUV luminosities.  
Therefore, any differences between the morphological distributions of these
two samples is likely to be the results of intrinsic differences in the samples
as opposed to selection effects.   However, the selection functions are not identical. 
Therefore we have selected those objects from the $z \sim 1.5$ sample with measurable
morphologies ($r_p > 0.3, \langle S/N \rangle \ge 2.5$) and artificially redshifted
these to $z \sim 4$.  The redshifted images were given the same exposure times and sky
background as the UDF $V+i$ images.  We find that $z \sim 1.5$ objects with isophotal 
surface brightnesses $< 23.6$ mag per sq arcsec in the GOODS $B$ images drop below the
UDF $V+i$ limiting surface brightness of $\sim 26.1$ mag per sq arcsec when artificially
redshifted because of the $(1+z)^4$ surface brightness dimming.  Therefore only
23 of the artificially redshifted objects have $\langle S/N \rangle \ge 2.5$. 
For these objects, the offsets in morphology between the original and redshifted images
are similar to those found in Equation 6.  The measured $r_p$ and isophotal magnitudes for
the artificially redshifted sample is shown in Figure 3. The UDF $z\sim4$ sample probes a
volume roughly four times that of the un-redshifted GOODS $z \sim 1.5$ sample, assuming
a limiting magnitude of $V+i \sim 27$ and $B \sim 25$ for each sample respectively.

\begin{figure}
\plotone{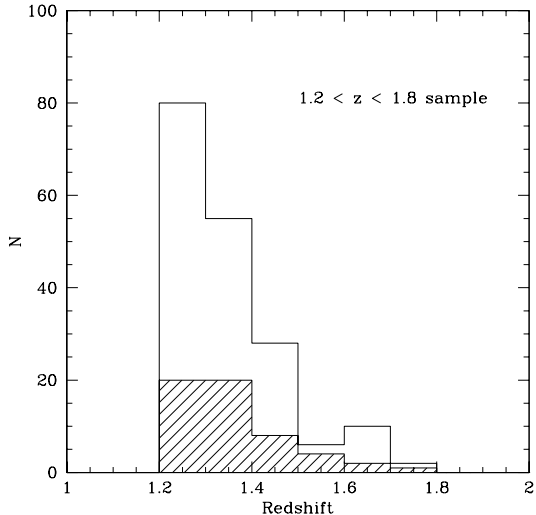}
\caption{Spectroscopic redshift histogram for GOODS $1.2 < z < 1.8$ 
sample (shaded histogram: 55 galaxies with $\langle S/N \rangle \ge 2.5$, $r_p \ge 0.3 \arcsec$
selected for the morphological analysis) . }
\end{figure}

\section{GALAXY MORPHOLOGIES at $z \sim$ 1.5 and $z \sim$ 4}
In Figures 5-7, we show the $G$, $M_{20}$, and $C$ values for all galaxies with 
$\langle S/N \rangle \ge 2.5$ and $r_p \ge 0.3$\arcsec (small crosses), as
well as the artificially redshifted sample.  
We find that the $z \sim 4$  LBGs have median G $\sim 0.58$, $M_{20} \sim -1.6$ and $C \sim 3.8$.  
We note that Ferguson et al. (2004) found similar concentration values for the
ACS $z$-band observations of GOOD LBGs, but that Conselice et al. 2004 and LPM04 
reported significantly lower concentration values for the observed 
near-infrared morphologies of $2 < z < 3.5$ galaxies in the
Hubble Deep Field North. We suspect that the larger PSF of the NICMOS images has biased
the measured concentrations to lower values.  The mean $G$ and $M_{20}$ values are very
similar to those found by LPM04 for the rest-frame $u$ and $B$ morphologies of
$z \sim 2$ and $z \sim 3$ galaxies in the Hubble Deep Field North.

A series of two-dimensional Kolmogorov-Smirnov (K-S) tests (Fasano \& Franceschini
1987) give a less than 37\% probability that bright GOODS LBGs and the fainter UDF LBGs 
are drawn from the different distributions in $G$, $M_{20}$ and $C$. 
The $z \sim 1.5$ sample's median morphological values are slightly different,
with median $G \sim 0.55$ (with a tail to lower $G$), $M_{20} \sim -1.5$, and
$C \sim 3.3$.  These median values do not change when the $z \sim 1.5$ sample is
artificially redshifted to $z = 4$.  The K-S statistical probability that the morphological distributions
of the artificially-redshifted $z \sim 1.5 \rightarrow 4$ sample is different from 
the UDF $z \sim 4$ LBG sample is greater than 91.5\%.

\begin{figure*}
\plotone{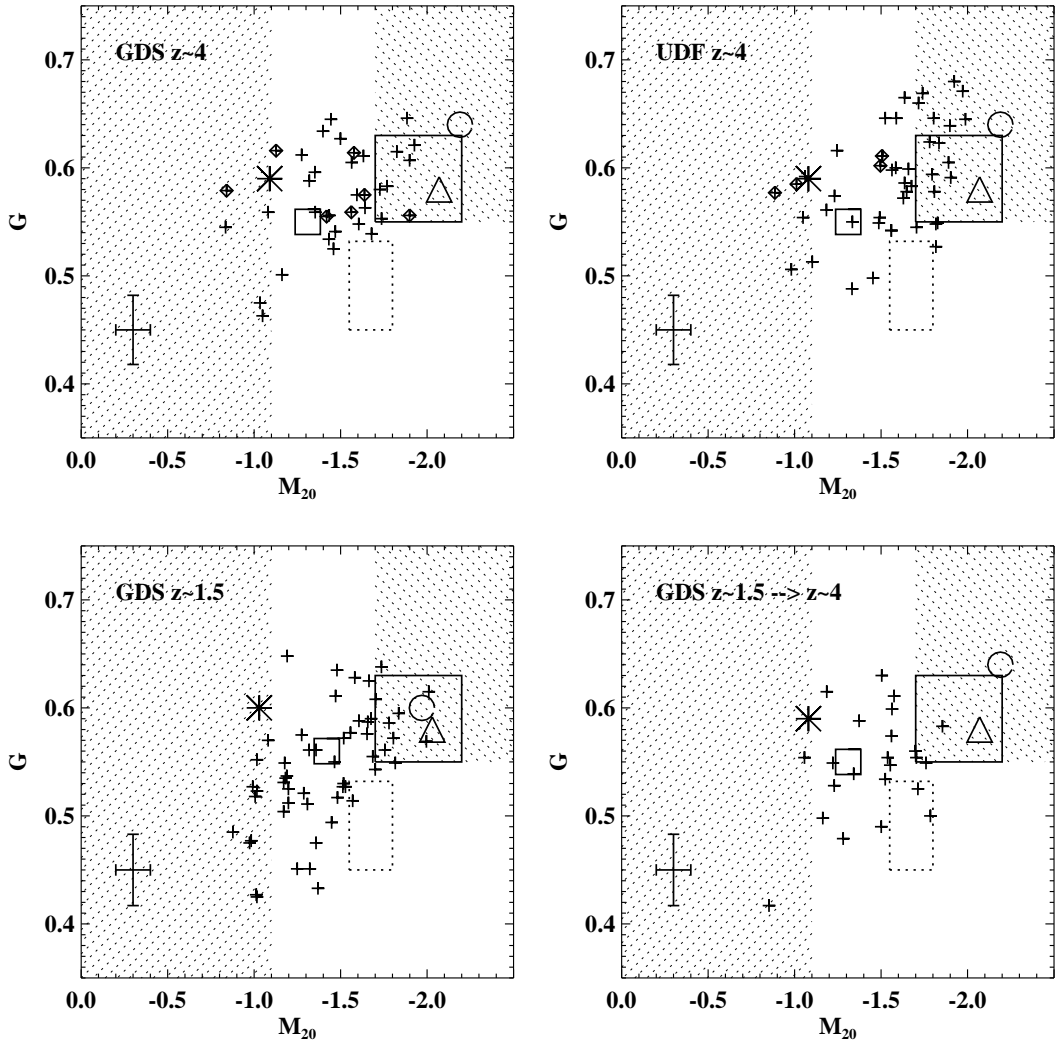}
\caption{$G$ v. $M_{20}$ for GOODS and UDF $z \sim 4$, and GOODS 
$z\sim 1.5$ and artificially redshifted $z \sim 1.5 \rightarrow 4$ 
 $\langle S/N \rangle \ge 2.5$, $r_p \ge 0.3 \arcsec$ samples (pluses).
LBGs with modified segmentation maps (\S3.1) have been marked with diamonds.
The upper and lower dotted boxes are where simulated face-on bulges
and disks $G$ and $M_{20}$ values lie, respectively. 
The FUV morphologies of artificially redshifted local galaxies are also
plotted (NGC1399:circle, NGC1068:triangle, NGC2403:square, NGC520:star).  
Objects with $M_{20} < -1.1$ possess double or multiple bright nuclei 
and are likely to be mergers (left shaded region).  Objects with $M_{20} < -1.8$
and $G > 0.57$ are bulge-dominated (right shaded region).
The error-bar is typical for $\langle S/N \rangle = 2.5$ galaxy.}
\end{figure*}

\begin{figure*}
\plotone{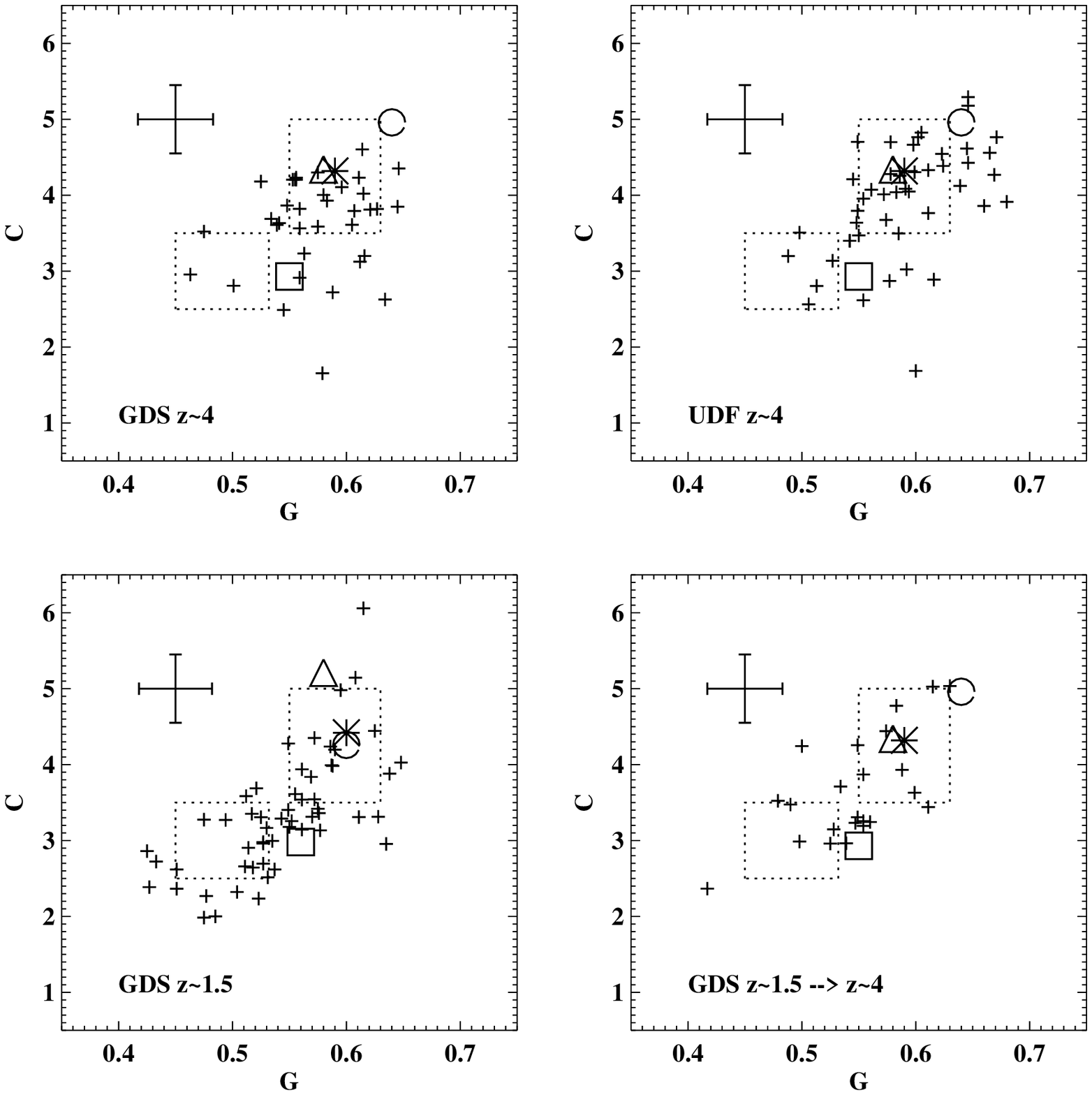}
\caption{$C$ v. $G$ for GOODS and UDF $z \sim 4$, GOODS $z\sim 1.5$ and artificially
redshifted $z \sim 1.5 \rightarrow 4$ 
$\langle S/N \rangle \ge 2.5$, $r_p \ge 0.3 \arcsec$ samples.
The symbols are the same as for Figure 5.}
\end{figure*}

\begin{figure*}
\plotone{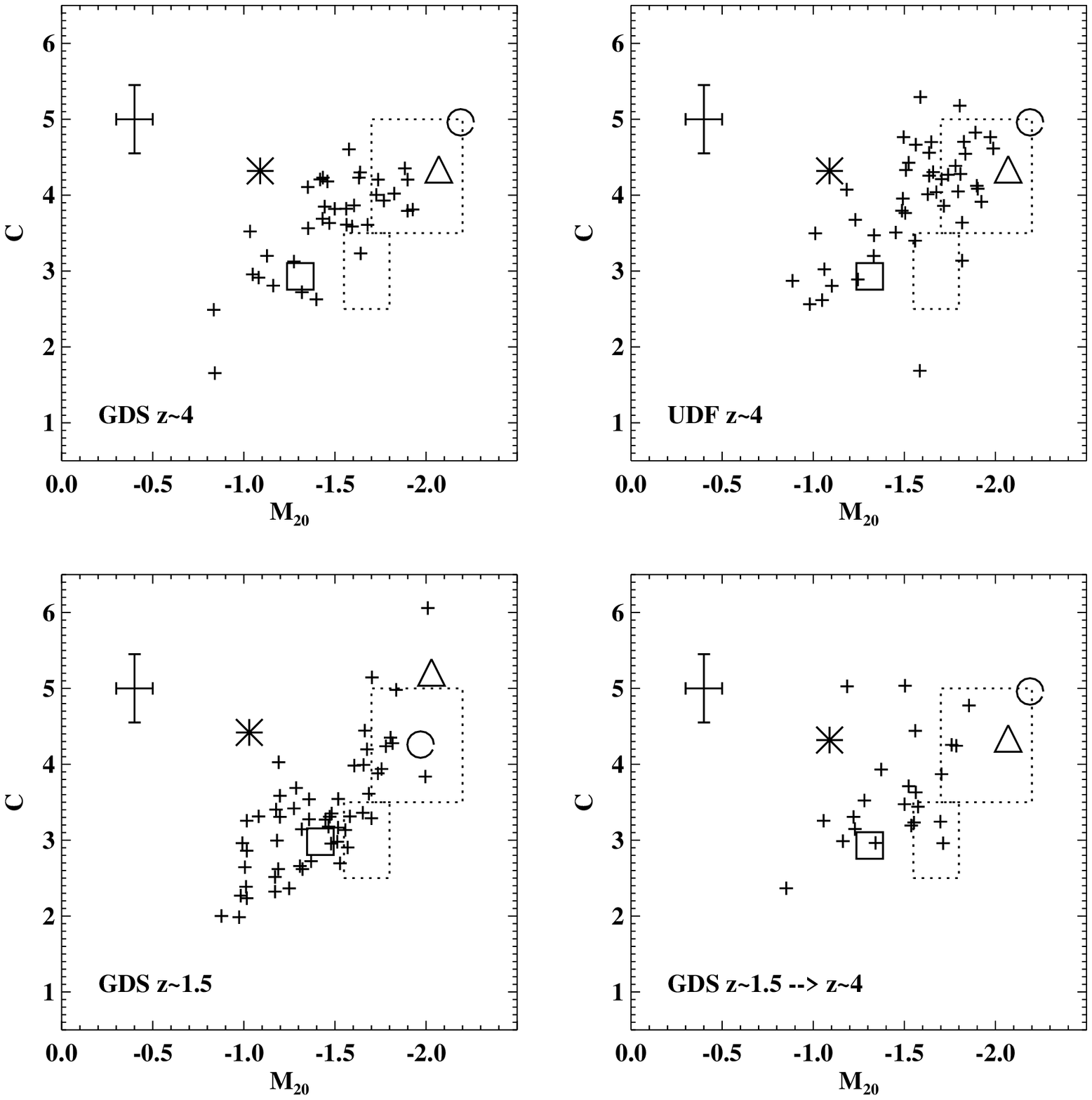}
\caption{$C$ v. $M_{20}$ for GOODS and UDF $z \sim 4$, GOODS $z\sim 1.5$ and
artificially redshifted $z \sim 1.5 \rightarrow 4$ 
$\langle S/N \rangle \ge 2.5$, $r_p \ge 0.3 \arcsec$ samples.
The symbols are the same as for Figure 5. }
\end{figure*}

The postage stamps for a subset of galaxies in both redshift
ranges are shown in Figures 8-10.  In all samples, galaxies with 
$M_{20} \geq -1.1$ have well-separated
double or multiple bright nuclei.  Galaxies with $M_{20}
\leq -1.6$ are relatively smooth with a single nucleus.
Objects with intermediate $M_{20}$ values can be irregular in
appearance, often with a bright nucleus with tidal tail or fainter
knots. Galaxies with $G > 0.6$ have very bright nuclei, 
and include both smooth ``bulge''-like objects and close pairs. 
In the lower-redshift sample, we observe a number of 
galaxies with $G < 0.55$, $M_{20} > -1.5$, $C < 3$ . These tend to have 
extended, lower-surface brightness star-forming regions,
reminiscent of local disk galaxies (Figure 10). 

\begin{figure*}
\plotone{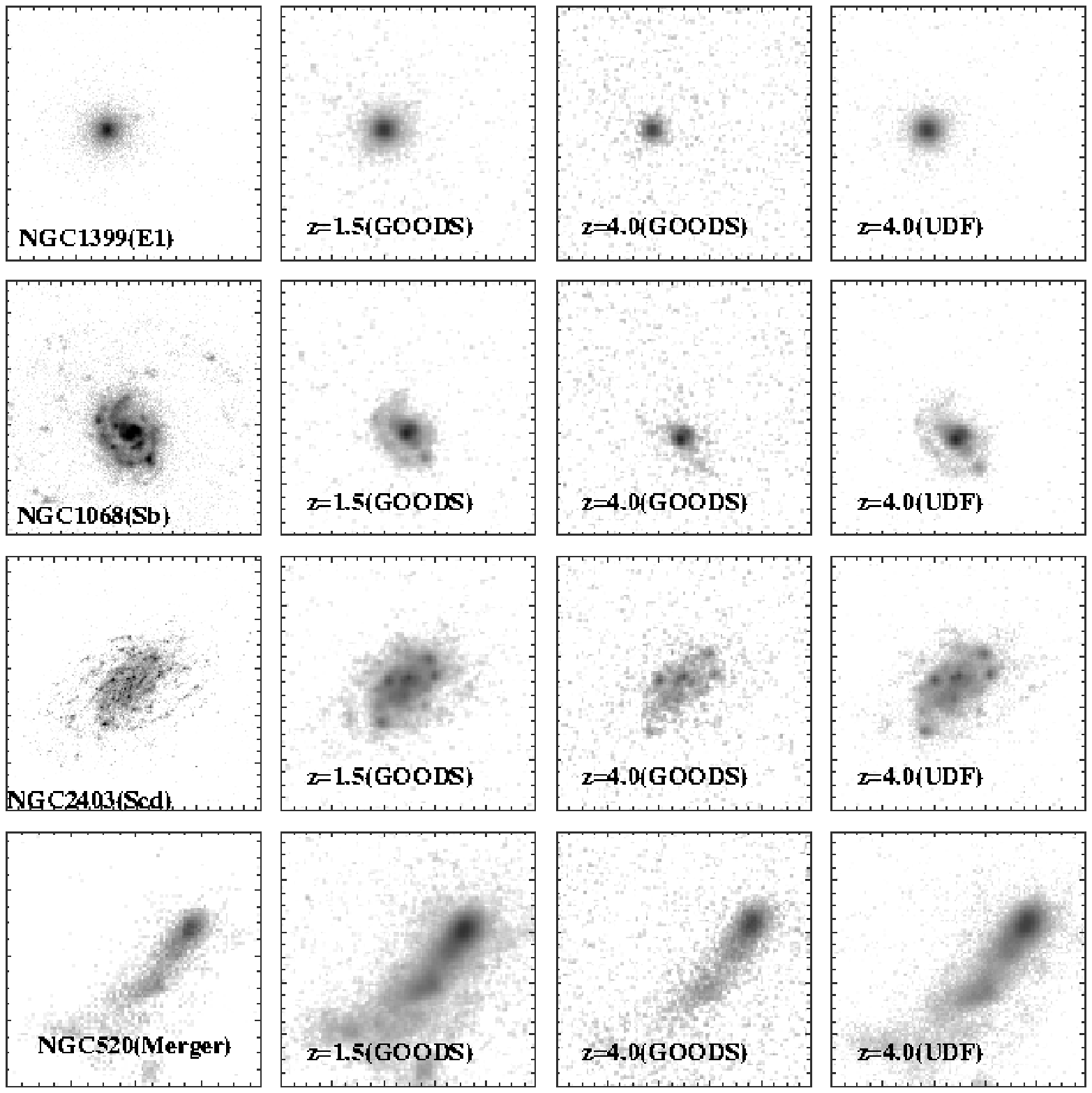}
\caption{GALEX FUV images of four nearby galaxies have been artificially
redshifted to $z=1.5$ and $z=4$, brightened to $M_{FUV} = -21$,
and simulated to match the GOODS $B$, GOODS and UDF $V+i$
observations.}
\end{figure*}

\subsection{FUV Morphological Classification}
Although we are able to robustly  measure the rest-frame FUV
morphologies of our $z \sim 1.5$ and $\sim 4$ samples, 
we cannot simply apply the quantitative classification criteria
that was originally developed for the optical morphologies of local
galaxies.  Young star-forming populations completely dominate the
rest-frame FUV morphologies of local galaxies, while old or dust-enshrouded
stellar populations all but disappear.  Thus the FUV morphologies of
normal galaxies appear significantly more disturbed than their optical
morphologies (Giavalisco et al. 1996, Hibbard and Vacca 1997; Goldader et al. 2002). 
In order to re-calibrate our merger classification criteria, we have 
examined the FUV morphologies of four bright well-resolved galaxies observed
by the GALEX Nearby Galaxy Survey.   While a detailed study of a much larger
sample of galaxies is needed, our preliminary work here in combination with
our bulge and disk simulations provides a check on our revised
FUV morphology classifications.

The GALEX FUV images of NGC1399 (E1), NGC1068 (Sb), NGC2403 (Scd), and NGC520 (Arp 157, 
merger) have been artificially redshifted to $z=1.5$ and $z=4$, 
convolved with the ACS PSF simulated using TinyTim, 
and added to the GOODS $B$, $V+i$, and UDF $V+i$ sky background (Figure 8).  
Because the absolute $M_{FUV}$ of the local galaxies are typically $\ge -16$
and would be undetectable at $z > 1.5$, the artificially redshifted galaxies 
have been brightened to $M_{FUV} = -21$.   We have not applied any K-corrections
to the images. We have measured the
morphologies of the artificially redshifted galaxies as they appear at
$z = 1.5$ and $z = 4$ and plotted their $G$, $M_{20}$, and $C$ values in
Figure 5-7.  The elliptical NGC1399 (circle) has $G \ge 0.6$, $M_{20} \le -2.0$
and $C \ge 4.3$, in rough agreement with our $r^{1/4}$ bulge simulations (upper
right-hand box).
The Sb NGC1068 (triangle) has $G \sim 0.58$, $M_{20} \sim -2.0$, and $C > 4.3$, 
which also agrees with our bulge simulations.   The Scd NGC2403 (square) has
$G \sim 0.55$, $M_{20} \sim -1.4$, and $C \sim 2.9$.  This gives a slightly
higher $G$ and $M_{20}$ values than our smooth exponential disk simulations
(lower left-hand box), as might be expected for a real spiral with bright star-forming knots.
The merger NGC520 (star) has $G \sim 0.59$, $M_{20} \sim -1.0$, and $C \sim 4.3$. 
NGC520 has been classified as an intermediate stage merger of two gas-rich
spirals (Hibbard \& van Gorkum 1996), and only one of its nuclei is visible in the FUV. 

Based on the redshifted GALEX images, our bulge and disk simulations, and
visual inspection, we have classified the $z \sim 1.5$ and $z \sim 4$ FUV
morphologies into three types. Major-merger candidates such as close pairs 
and/or objects with multiple bright nuclei or bright tidal tails 
have $M_{20} \ge -1.1$ (left shaded region of Figure 5).  Bulge-dominated
systems (E-Sb) are those with $G > 0.57$ and $M_{20} < -1.7$ (upper right shaded
region of Figure 5).  Galaxies which do not match any of
these criteria are called ``transition'' objects. We stress that this
initial classification scheme is more conservative than that of LPM04 because
of the lack of a large calibration sample of FUV galaxy images  and
will only identify major mergers in their most morphologically-disturbed stage.
We expect that some of the transition objects will be minor-mergers and post mergers, 
as well as late-type star-forming disks which appear knotty in the rest-frame FUV.
We find that $C$-based classifications often disagree with those
based on  $G-M_{20}$, as $C$ is less robust at small sizes and is 
less sensitive to substructure (Figures 6 and 7).

\begin{figure*}
\plotone{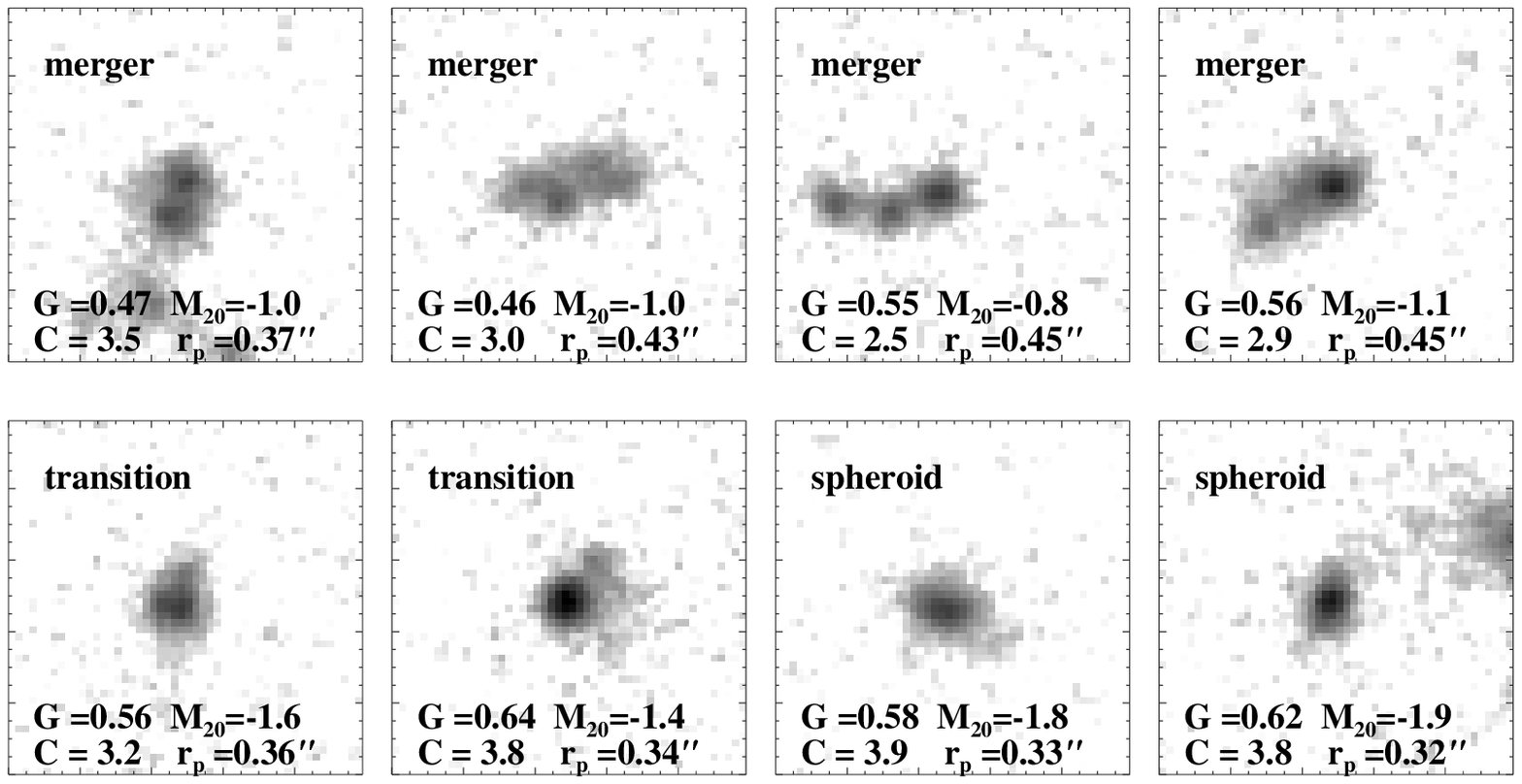}
\caption{$V+i$ 1.5\arcsec\ $\times$ 1.5\arcsec\ images for a subset 
of GOODS $z \sim 4$ LBGs. Classifications are based on Figure 5.}
\end{figure*}

\begin{figure*}
\plotone{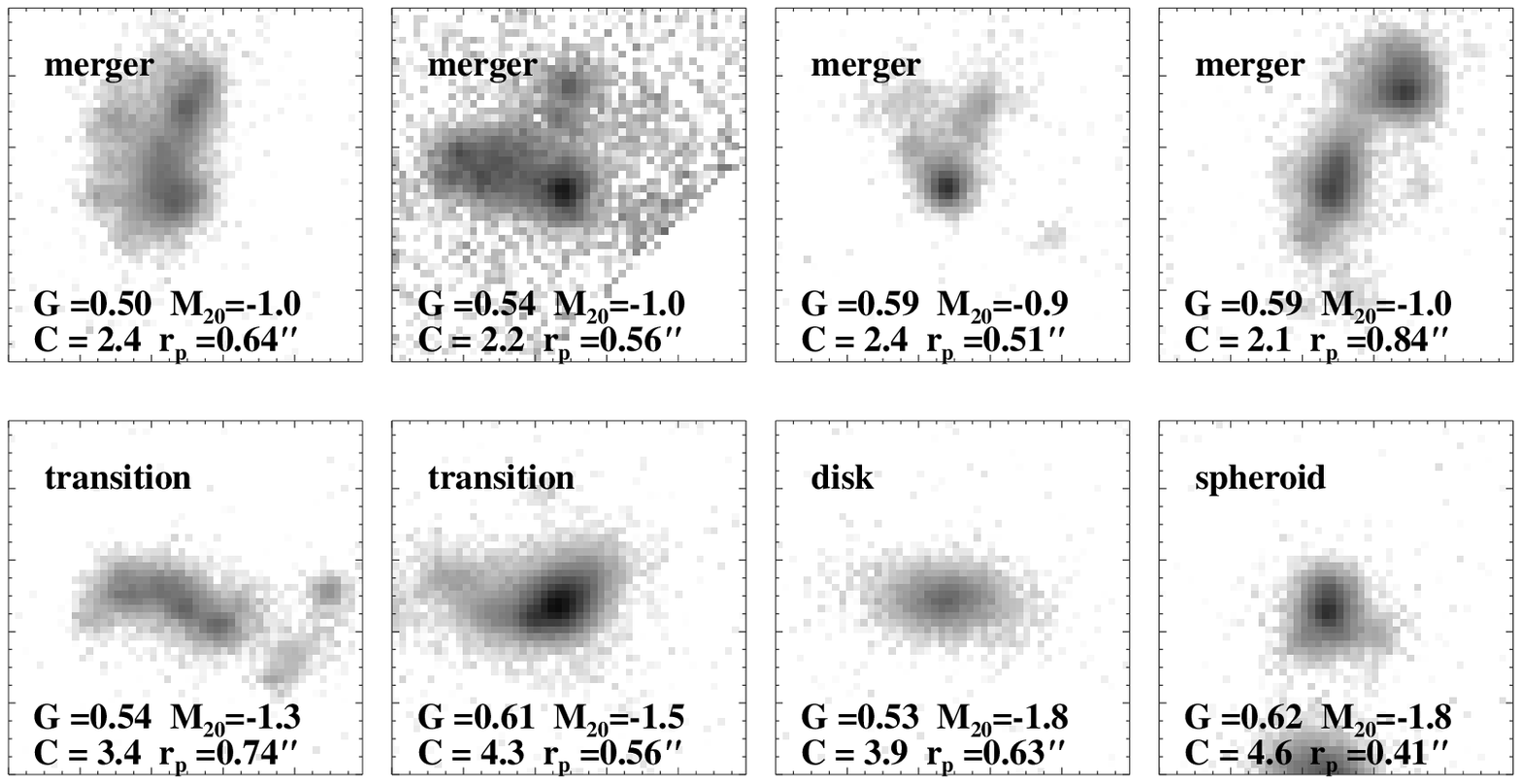}
\caption{$V+i$ 1.5\arcsec\ $\times$ 1.5\arcsec\ images for a subset 
of UDF $z \sim 4$ LBGs. Classifications are based on Figure 5.}
\end{figure*}

\begin{figure*}
\plotone{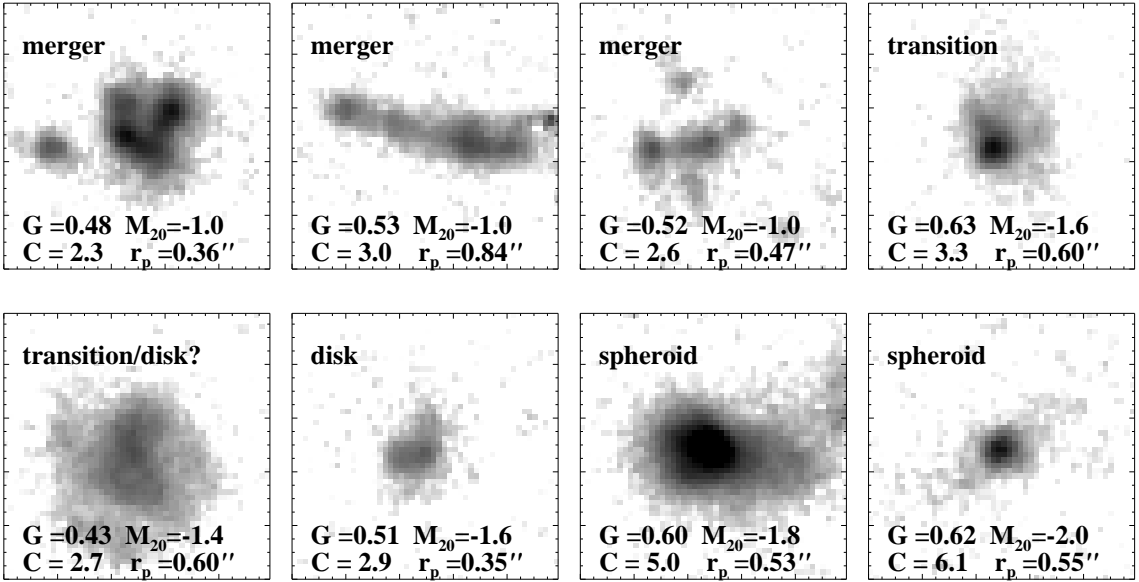}
\caption{$B$ 1.5\arcsec\ $\times$ 1.5\arcsec\ images for a subset 
of GOODS $z \sim 1.5$ galaxies. Classifications are based on Figure 5.}
\end{figure*}

\begin{deluxetable*}{llrlcrlcrl}
\tabletypesize{\scriptsize}
\tablecolumns{10}
\tablewidth{0pc}
\tablecaption{Morphology Distributions}
\tablehead{ 
\colhead{} & \colhead{} & \multicolumn{2}{c}{$M_{FUV} < -21$}  & \colhead{}
& \multicolumn{2}{c}{$-21 < M_{FUV} < -20$} & \colhead{} &\multicolumn{2}{c}{$-20 < M_{FUV} < -19$} \\
\cline{3-4} \cline{6-7} \cline{9-10} \\
\colhead{Sample} &  \colhead{Type}  & \colhead{N$_{obs}$}   & \colhead{Fraction}  &\colhead{} 
& \colhead{N$_{obs}$}   & \colhead{Fraction}  & \colhead{} & \colhead{N$_{obs}$}   & \colhead{Fraction} }
\startdata
GOODS $z \sim 4$ &  Total      & 11   & \nodata      &   & 25   &\nodata      &  & \nodata  &\nodata \\
                 &  Mergers    &  2   & 0.47 (0.18)  &   &  3   & 0.21 (0.12) &  & \nodata  &\nodata \\
                 &  Bulges     &  3   & 0.19 (0.27)  &   &  8   & 0.23 (0.32) &  & \nodata  &\nodata \\
                 &  Other      &  6   & 0.34 (0.55)  &   & 14   & 0.56 (0.56) &  & \nodata  &\nodata \\
\cutinhead{}
UDF $z \sim 4$   &  Total      &  5   & \nodata      &   & 19   &\nodata      & & 22        &\nodata \\
                 &  Mergers    &  3   & (0.6)       &   &  1   & (0.05)      & &  1         & (0.05)  \\
                 &  Bulges     &  1   & (0.2)       &   &  7   & (0.37)      & &  11         & (0.50)  \\
                 &  Other      &  1   & (0.2)       &   &  11   & (0.58)      & & 10         & (0.45)  \\
\cutinhead{}
GOODS $z \sim 1.5$ & Total     & 3    & \nodata      &   & 23   & \nodata     & & 28        &\nodata \\ 
                 &  Mergers    & 1    & (0.33)       &   &  4   & (0.17)      & &  5         &(0.18)  \\
                 &  Bulges     & 2    & (0.66)       &   &  4   & (0.17)      & &  8         &(0.30)  \\
                 &  Other      & 0    & (0.00)       &   & 15   & (0.65)      & & 16          &(0.57)\\
\cutinhead{}
GOODS $z \sim 1.5$ & Total     & 2    & \nodata      &   & 8   & \nodata     & & 13        &\nodata \\ 
$\rightarrow z= 4$    &  Mergers    & 1    & (0.5)       &   &  1   & (0.125)      & &  0         &(0.0)  \\
                 &  Bulges     & 0    & (0.0)       &   &  1   & (0.125)      & &  2         &(0.15)  \\
                 &  Other      & 1    & (0.5)       &   &  6   & (0.75)      & & 11          &(0.85)
\enddata
\tablecomments{The fractions in parentheses are the number of galaxies in the each morphological bin divided
by the total measured objects in each sample and have not been corrected for incompleteness.}
\end{deluxetable*}

\subsection{Merger and Spheroid Fractions at $z \sim 1.5$ and $z \sim 4$}
In Table 1, we give the observed merger and bulge fractions for our UDF $z \sim 4$ 
and GOODS $z\sim 4$ LBGs, the GOODS $z \sim 1.5$ star-forming galaxies, and the
artificially redshifted sample. At $M_{FUV} < -20$,  the observed fraction of merger 
candidates is similar for all samples ($\sim$ 10\% - 20\%).  
In Figure 3, we showed that the GOODS and the UDF LBG samples 
probe different parts of the luminosity function, while the UDF and $z \sim 1.5$
samples are better matched.  In this section, we will use the combined GOODS and UDF
LBG samples to constrain the morphological distribution of $z\sim4$ star-forming galaxies,
but use only the UDF $z\sim4$ and $z\sim1.5$ samples to constrain any evolution with redshift.

If we classify all objects with $M_{20} \le -1.1$
as on-going major merger candidates, the observed major merger fraction at $z \sim 4$ is
$\sim$ 14\% for GOODS Bdrops (5/36) and $\sim$ 10\% for UDF Bdrops (5/46). 
However, we have not corrected this fraction for incompleteness.  
In order to do a rough incompleteness correction, 
we have visually inspected the LBGs with $\langle S/N \rangle < 2.5$ and
$V+i$ magnitudes brighter than 25.  These objects have $r_p > 0.5$\arcsec and were excluded
from our original sample because of their low surface brightness and $\langle S/N \rangle$.  
Fifteen out of the 25 bright excluded objects were visually classified as disturbed, therefore it
is possible that we are missing a significant fraction of large, bright but lower-surface
brightness merger candidates.  Four of these objects were visually classified as
bulge-dominated.   We have also excluded a large number of low-surface
brightness GOODS LBGs at $25 < V+i < 26$. Examination of the UDF LBGs at the
same magnitude and sizes indicates that $\sim 22$\% of the excluded GOODS LBGs with
$25 < V+i < 26$ are likely to be major merger candidates and $\sim 22\%$ 
are likely to be bulge-dominated.  If we correct the GOODS LBG merger fraction for
these rough incompleteness estimates, we find that average fraction of major-merger candidates
is $\sim 25 \% $ for all LBGs with $V+i < 26$, $r_p > 0.3$\arcsec.  If we consider
only the brightest LBGs with $V+i < 25$, $r_p > 0.3$\arcsec, the major-merger
fraction could be as high as $\sim 50$\% once incompleteness effects are included.
However, our corrections are highly uncertain, we do not account for LBGs which
are undetected, and the number of bright LBG is small (16).  
Nevertheless, these may be considered rough upper limits
to the morphologically-determined merger fraction for star-forming galaxies.

It is unclear whether our major-merger fraction for $z \sim 4$ LBGs
is consistent with previous high-redshift quantitative morphology 
studies based on the rotational asymmetry parameter.  
Conselice et al. (2003) found a 50\% merger fraction for $M_B < -21$ galaxies 
at $z \sim 2-3$ based on a sample of 10 objects in the HDFN. 
If we assume $(FUV-B) \sim 1$, then our merger fraction estimate of 25\% 
is about a factor of two lower for the same magnitude range ($M_{FUV} < -20$) 
based on a sample of 60 LBGs in GOODS and UDF.  
However, at $M_{FUV} < -21$ ($M_B < -22$) our visual classification suggests 
that $\sim 50\% $ are disturbed objects, 
so there may be a trend towards higher merger fractions at brighter luminosities. 

We find that a significant fraction of $z \sim 4$ objects have
morphologies consistent with spheroidal morphologies:  
11/36 of the GOODS LBGs and 19/46 of the UDF LBGs lie within the spheroid criteria
in $G-M_{20}$ space.  Brighter than $M_{FUV} \sim -20$, roughly 30\% of the GOODS
and UDF galaxies appear to be bulges. Many of the UDF spheroids are fainter than $M_{FUV} 
= -20$ (11/22). This is in contrast to the $M_{FUV} \ge -20$  $z \sim 1.5$ galaxies, 
of which only 2/15 of the artificially redshifted sample are bulge-like.

The remaining $\sim 50\%$ of the galaxies in the 
high-redshift samples have higher $M_{20}$ and $G$ 
than expected for smooth disks or bulges but lower $M_{20}$ values than
the major-merger candidates. Some of these objects lie close 
in $G-M_{20}$ to the artificially-redshifted
FUV image of late-type spiral NGC2403, therefore many 
may be star-forming disks.  In the local universe at optical wavelengths, 
higher $M_{20}$ and $G$ values are indicative of merger signatures
such as very bright nuclei and tidal tails, and it is also possible that
a fraction of the transitional objects are minor-mergers or post-merger
systems which are disturbed but do not possess bright multiple nuclei
and high $M_{20}$ values.  

The $z\sim1.5$ objects are more likely to fall into this transitional
class than the UDF LBGs (60\% of the $z \sim 1.5$ sample and 70\% of the
artificially redshift sample).   Also, 
the $G$ distributions of the transitional objects are slightly
different at the two epochs:  less than 5\% (1/46) of the observed UDF $z \sim 4$ 
sample has $G \le 0.5$, as opposed $\sim 20\%$ at lower redshift ( 10/55 of the 
$z \sim 1.5$ sample and 5/23 of the artificially 
redshifted $z \sim 1.5 \rightarrow 4$ sample). 
Low $G$ values usually indicate disk-like galaxies with more
uniform surface brightness profiles.  Is the lack of low $G$ galaxies
at high redshift the result of an observational bias?  We clearly select against large 
disks because of their low-surface brightnesses.  
However the lower-redshift sample has low $G$ objects 
with sizes, total magnitudes, and effective surface brightnesses 
which should have been detectable in the UDF (Fig 3).  
Our artificial redshift simulations suggest that the detected fraction of low $G$ objects
should not change due to redshift-dependent selection effects. 

The slightly lower Gini coefficients, concentrations, 
and second-order moments at $z \sim 1.5$ may be partially a 
result of larger dust obscuration.
Dust may have significant effect on the observed morphologies in the FUV than
at longer wavelengths. In a study of
local ULIRGs with {\it HST} STIS images in the
FUV/NUV and WFPC2 and NICMOS images at optical and near-infrared
wavelength, Goldader et al. (2002) found than many star-forming knots
were obscured in the FUV/NUV, resulting in substantial changes in the
observed morphologies and light distributions.  The $z \sim 4$
Lyman break galaxies may not suffer from large dust extinction 
because Lyman break technique automatically selects blue objects with
low to moderate extinctions; 
however, the $z \sim 1.5$ sample is likely to 
possess dust and have higher internal color
dispersions. In a study of the internal FUV-optical color dispersions
of Hubble Deep Field North $z \sim 2.3$ and $z \sim 1$ samples,
Papovich et al. (2005) found that the lower redshift sample 
had significantly more internal color dispersion than the high redshift
sample, a result of either patchier dust distributions or
larger spatial dispersions in the age of the stellar populations.

\subsection{ $FUV-B$ Colors and UV Slope}
The southern GOODS field (CDFS) has deep ground-based near-infrared imaging
in $J$, $H$, $K_s$ with the ISAAC camera at the VLT (Vandame et al 2004).  
The GOODS ACS images were degraded to the ISAAC image resolution, 
and matched aperture photometry was obtained for
the CDFS/UDF galaxies (Dahlen et al. 2005).  Eight GOODS-S LBGs and twelve 
UDF LBGs (one of which is also in the ISAAC-detected GOODS LBG sample) 
from our $\langle S/N \rangle \ge 2.5$, $r_p \ge 0.3~\arcsec\ $ were detected in
the ISAAC $K_s$ image and 16 GOODS-S $z \sim 1.5$ galaxies were detected in 
the ISAAC $J$ image.   We have examined the rest frame $B$ isophotal magnitudes ($K$ and $J$
for $z \sim 4$ and $z \sim 1.5$ samples), rest-frame $FUV - B$ isophotal colors, 
and rest-frame $FUV-NUV$ isophotal colors as a function of galaxy morphology in Fig. 12  and 13.
At $z \sim 4$, the $z-K$ color corresponds to $\sim$ 1700\AA\ - 4400\AA, while at
$z \sim 1.5$, the $B-J$ color corresponds to $\sim$ 1740\AA\ - 4400\AA. 
We use the rest-frame $FUV-NUV$ colors as a proxy for 
the UV spectral slope $\beta$ ($F_{\lambda} \propto \lambda^{\beta}$, Calzetti et al. 1994), 
using the ISAAC-degraded $z-J$ isophotal colors ($\sim$ rest-frame 1700\AA\ - 2200\AA)
for the $z \sim 4$ galaxies and using the ACS $B-V$ isophotal colors ($\sim$ rest-frame 1740\AA\ - 2500 
\AA) for the $z \sim 1.5$ galaxies.

We find no strong correlation of rest-frame 4400\AA\ luminosity with 
FUV galaxy morphology for either sample (upper panels, Fig. 12 and 13).
We find that majority of $z \sim 4$ LBGs have $z-K < 1.5$. There are two red
LBGs with $i-K > 2.5$; such red rest-frame $FUV- B$ colors indicate strong 4000\AA\ breaks 
and relatively old stellar populations ($> 0.5 - 1$ Gyr), while the bluer colors 
of the majority of the $z \sim 4$ LBGs indicate
younger stellar populations. The blue LBGs show no obvious 
correlation of rest-frame $FUV- B$ color or UV slope ($z-J$ or rest-frame $FUV-NUV$) 
with $G$, $M_{20}$ or $C$.  
The two red LBGs with $i-k > 2.0$ have morphologies consistent with 
spheroids ($G > 0.5, M_{20} < -1.5$) and are also the brightest objects
in the $K$-band sample.   The $z \sim 1.5$ sample has similar $FUV-B$ colors
but show steeper UV slopes ($B-V$ or rest-frame $FUV-NUV$), 
implying either larger dust extinctions or older
ages.  There are no obvious color-morphology trends.  The reddest object in the 
$z \sim 1.5$ sample is also consistent with
a bulge-like morphology ($G =0.60, M_{20} = -1.7, C=5.1$).

\begin{figure*}
\plotone{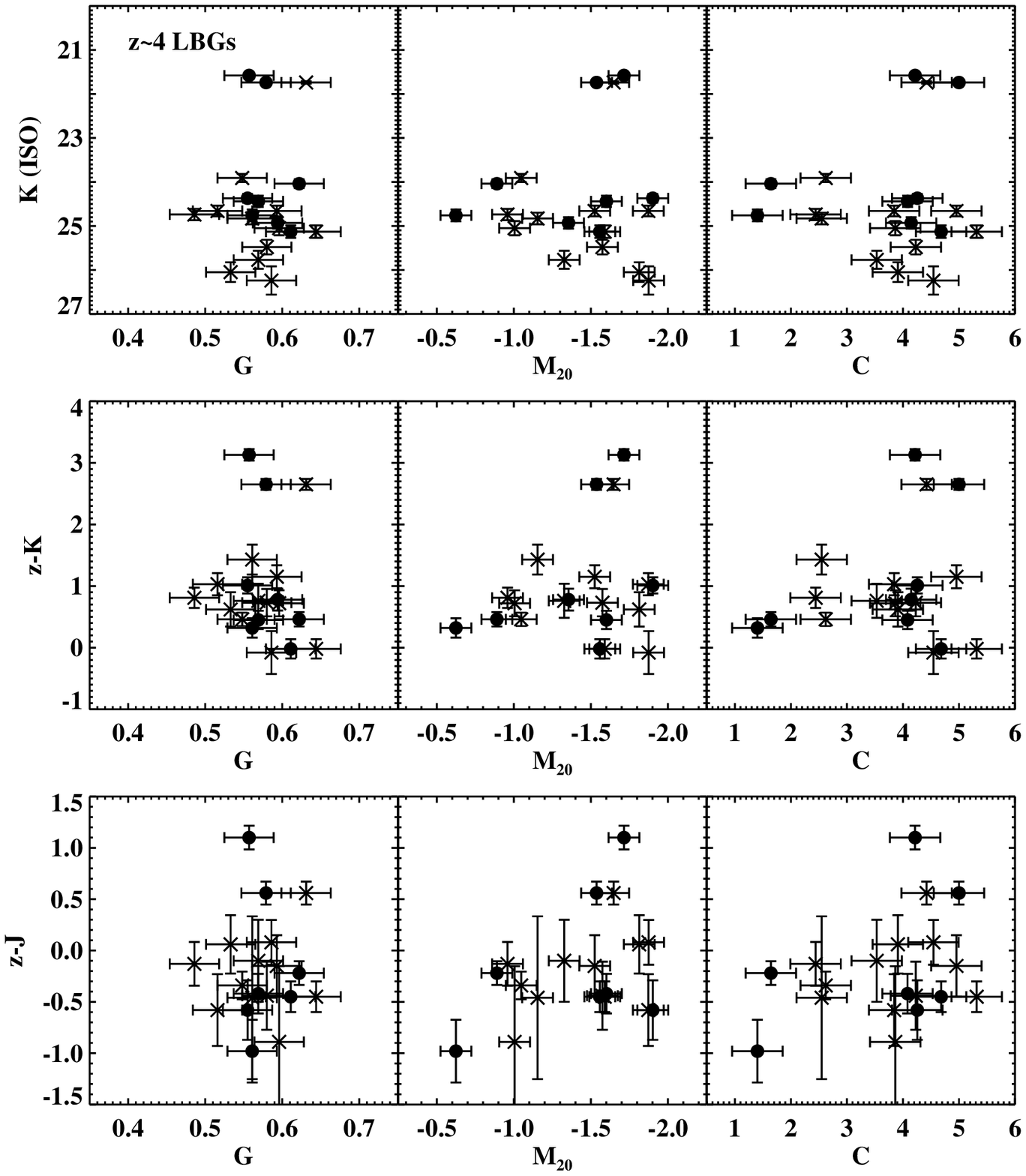}
\caption{Isophotal $K$ AB magnitudes, $z-K$ colors ($\sim$ rest-frame 
1700\AA\ - 4400\AA), and $z-J$ colors ($\sim$ rest-frame 1700\AA\ - 2200\AA)  
v. morphology for GOODS-South (crosses) and UDF (circles) $z \sim 4$ samples.}
\end{figure*}

\begin{figure*}
\plotone{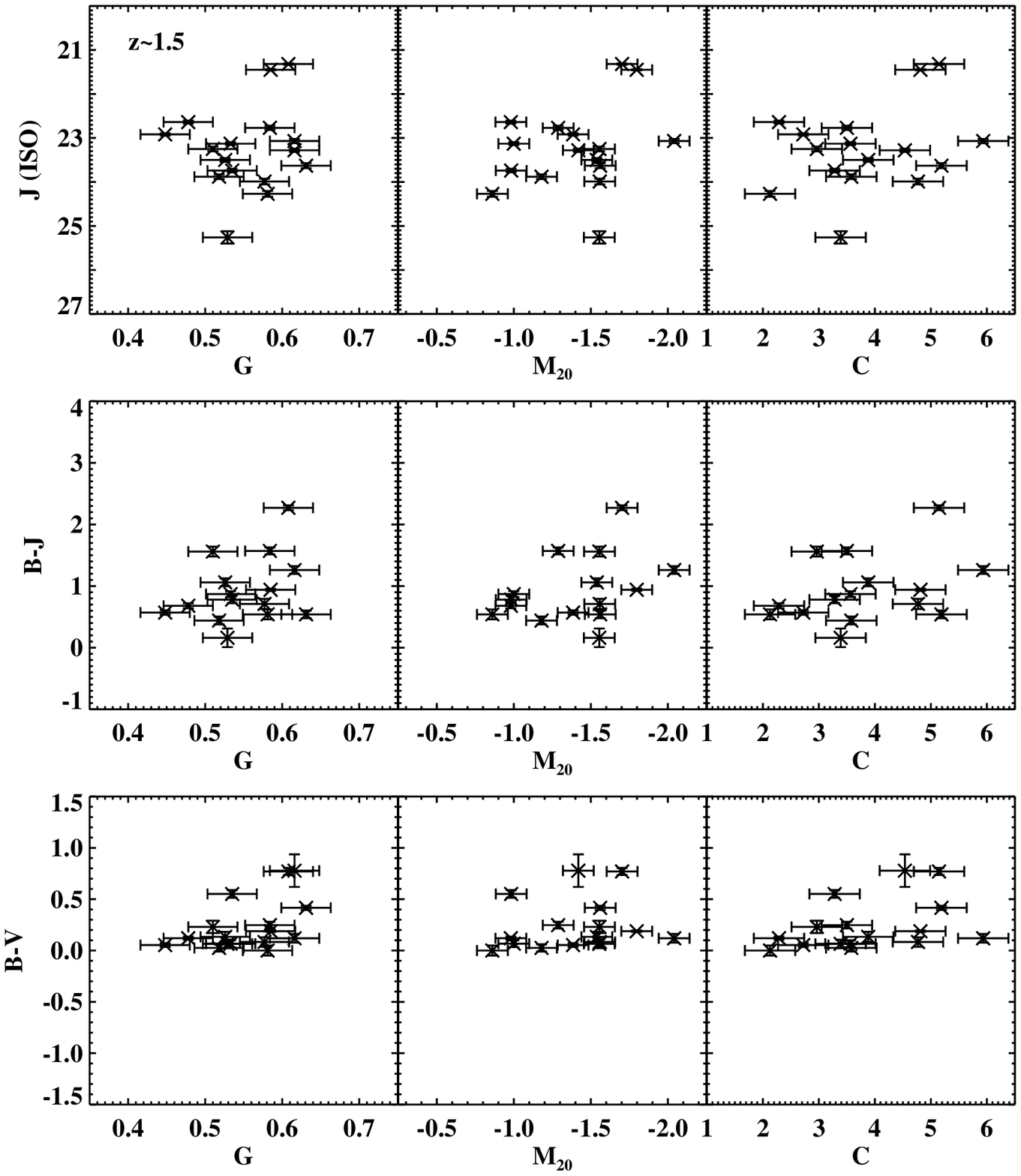}
\caption{Isophotal $J$ AB magnitudes, $B-J$ colors ($\sim$ rest-frame
1740\AA\ - 4400\AA), and $B-V$ colors ($\sim$ rest-frame 1740\AA\ - 2500\AA) 
v. morphology for $z \sim 1.5$ sample.}
\end{figure*}

\section{DISCUSSION}

We find no strong evolution in the rest-frame FUV morphologies of star-forming galaxies
between $z \sim 4$ and $z \sim 1.5$, with a difference in look-back times of roughly 3 
billion years. We see subtle changes in the fraction of spheroids and low $G$ disk-like objects, but
the fraction of objects with bright multiple nuclei is indistinguishable for the two samples.  
At first glance, this result is somewhat surprising
given the large time difference between the two observed epochs and the apparent increase
in dust extinction at later times. 

However, the dominant mode of star formation at both epochs is expected to be collisional starbursts
triggered minor-mergers -- 45\% of $\rho_*$ at $z \sim 1.5$ and 70\% of $\rho_*$ at $z \sim 4$ (SPF01).   
Therefore many FUV-bright galaxies at both epochs are likely to be minor mergers which
exhibit a dominant bright nucleus with a disturbed morphology.  In fact, this is what we find
for both our $z \sim 1.5$ and $z \sim 4$ samples.  Over half of the galaxies have been
classified as ``transition'' objects between the spheroidal morphologies and the
double-nucleated major-merger candidates. While it is likely that some of these transition galaxies 
are post-major mergers or knotty star-forming disks,  a significant fraction are likely to
be minor mergers.

We also find very similar fractions of major merger candidates at both epochs. 
Our UDF $z \sim 4$ major-major candidate fraction of observed star-forming galaxies 
is $\sim 10\%$ for $M_{FUV} < -19$, 
while the $z \sim 1.5$ major-merger candidate fraction is $\sim 10-20\%$ over the same
luminosity range.   We caution that these observed fractions are not ``merger rates''; 
in order to calculate a true merger rate, we must correct for the selection biases in our
sample against low-surface brightness or low star-formation objects as well as the
probability and timescales for observing mergers with $M_{20} < -1.1$. 
As we discussed in \S2, identifying non-star-forming galaxies at these epoch is challenging,
therefore we limit our analysis to star-forming galaxies only.    
The probability of observing a merging system during a period of disturbed morphology
will depend on a number of factors including the initial conditions of the merger
system (mass ratio, gas fraction, orbital parameters), the viewing angle, and 
the timescale of the merger.  A large suite of hydro-dynamical merger simulations
required to calculate this probability are currently being analyzed (Cox et al. 2005,
Jonsson et al. 2005, Lotz et al. in prep).  From our initial analysis of 
gas-rich major merger simulations,  we expect equal mass merging spirals
on a prograde orbit with a gas-fraction equal to 0.52 
to have $M_{20} > -1.1$ for $\sim 500$ Myr. 

The number of major mergers per halo per Gyr 
is predicted to be $\sim$ 15\% at $z \sim 1.5$ and $\sim$ 25\% at $z \sim 4$, where
the minimum halo circular velocities are $\ge$ 50 km s$^{-1}$
(Gottl\"ober, Klypin, \& Kravtsov 2001).  
If number of detected star-forming galaxies per typical halo
is $\sim 2$ and does not change strongly from $z \sim 4$ to $z \sim 1.5$, 
then our results are roughly consistent with the predicted major merger rate for halos.
Given that our samples are fairly small, and that we have not adjusted the observed major merger 
fractions for undetected non-star-forming galaxies, it is surprising 
that our numbers are so consistent with theory. Clearly a more detailed comparison
between the merger rates and morphology distribution between hierarchical models and
the data is needed that takes into account the many selection effects and biases.

At later times, quiescent star-formation should become increasingly important. SPF01 
predict that $\sim$ 40\% of star-formation density occurs in 
a quiescent (disk-star formation) mode at $z \sim 1.5$ as compared to 
$\sim$ 20\% at $z \sim 4$.  
This suggests that the fraction of FUV-bright disk galaxies
will increase at later epochs.  While we find relatively few ``smooth'' disks at either epoch,
there are more objects with $G < 0.5$ at $z \sim 1.5$ which are
consistent with low-surface brightness disks (see Figure 11) than at $z \sim 4$.  
These disk candidates make up $\sim 20\%$ of  the entire $z \sim 1.5$ 
sample and a third of the transition objects at that epoch.

Finally, the observed fraction of star-forming galaxies with spheroidal morphologies
is similar at both epochs. However, when the $z \sim 1.5$ sample is artificially redshifted
to $z \sim 4$ the fraction of spheroids in the faintest bin is significantly lower than
that observed in the UDF $z \sim 4$ sample. If these bulges are only formed as the end-stage of a major merger,
then similar fractions would be expected.  However, if minor-mergers and interactions
also produce blue bulges (Kannappan, Jansen, \& Barton 2004), 
then one might expect significantly higher rates of
star-forming spheroids at higher redshift given the increased importance of minor-mergers
at those times.  On the other hand, there may not be enough time before $z \sim 4$
for significant numbers of proto-bulges to form via minor-mergers, as the timescale for a minor merger 
is $\ge $ 2 Gyr (Cox 2004).

Thus the observed distributions of star-forming galaxy FUV morphologies and the weak evolution 
between $z \sim 1.5$ and $z \sim 4$ appears to be generally consistent with rates of major
and minor mergers and the importance of quiescent star-formation predicted at those
epochs by hierarchical galaxy evolution models.  However, we caution that we have largely ignored
the morphological effects of dust and the evolution of typical dust extinction with redshift.
The $FUV-NUV$ colors suggest that the $z \sim 1.5$ sample is more extincted than the
the LBGs, as might be expected if the galaxies are chemically evolving with time.  Strong dust
extinctions are likely to hide regions of intense star formation, and shift observed morphologies
to lower $G$ and possibly lower $C$ and $M_{20}$ values, as well as remove some heavily-extincted
objects from our samples altogether. Nevertheless, since the observed morphologies does change
significantly from $z \sim 4$ and $z \sim 1.5$, increased dust does not appear to dominate
the observed FUV morphologies for moderate extinctions. 
We also caution that our conclusions here only apply to 
galaxies which have sufficient star-formation to meet the LBG color-criteria at $z \sim 4$ 
and to have emission-line determined redshifts at $z \sim 1.5$.  We are most certainly missing
galaxies at both epochs with no or strongly enshrouded star-formation. 

\section{SUMMARY}

1) We find no significant difference between the rest-frame FUV morphological distributions
of GOODS and UDF $z \sim 4$ LBGs samples.  
Lyman break galaxies with morphologies consistent with on-going major mergers make
up $\sim 10-25\%$ of the $M_{FUV} < -20$ GOODS and UDF selected samples.  A significant fraction 
of the LBGs appear to have relatively undisturbed bulge-like FUV morphologies ($\sim 30\%$).
The remaining $\sim 50\%$ have $G$ and $M_{20}$ values higher than expected for
smooth exponential disks or bulges but do not possess double nuclei, and could be
a mixture of star-forming disk, minor-mergers, and post-merger systems.
The observed major-merger fraction and large number of minor-merger candidates
are consistent with current numerical (Gottl\"ober et al. 2001) and semi-analytic
model predictions (Somerville et al. 2001) for high redshift galaxies.

2) The star-forming $z \sim 1.5$ sample has rest-frame FUV morphologies which are not
significantly different from the UDF $z \sim 4$ galaxies, with similar fractions of
major-merger candidates.  The lower redshift sample
does possess slightly more objects with lower $G$ and slightly lower $C$ and 
higher $M_{20}$ values, implying more extended, lower surface 
brightness disk-like star-forming regions, as well as a smaller fraction of
faint spheroids.  This is also roughly consistent with the
major-merger rates and the increased importance of
quiescent (disk) star-formation at $z \sim 1.5$ predicted by current hierarchical
models. 

3) The two LBGs with rest-frame $FUV-B$ colors consistent with strong
4000\AA\ breaks and $> 500$ Myr stellar populations have bulge-like
morphologies, as does the reddest $z \sim 1.5$ galaxy.  The UV slopes
of the lower redshift sample are significantly steeper than those
of the $z \sim 4$ LBGs.  Although dust extinction may be more
important in the $z \sim 1.5$ samples, the apparent lack of
evolution in FUV morphology since $z \sim 4$ suggests that moderate
extinctions do not strongly bias the observed FUV $G$ and $M_{20}$ values.
High-resolution rest-frame optical morphologies for large samples of high
redshift objects are needed to constrain the morphologies of 
galaxies which are heavily dust-extincted or not forming stars. 

We thank the anonymous referee for his/her suggestions which improved
this paper. We would like to acknowledge the GOODS team for the ISAAC-ACS matched CDFS photometric catalog,
and the ESO/GOODS, the TKRS, and VIRMOS teams for making their redshift catalogs publicly
available. We would also like to thank C. Conselice, S. M. Fall, C. Papovich, S. Ravindranath,
and R. Somerville for their comments on this manuscript. 
Support for this work was provided by NASA grant NAG5-11513 and NSF Grant AST-0205738.
The GALEX data presented in this paper were obtained from the Multi-mission 
Archive at the Space Telescope Science Institute (MAST). STScI 
is operated by the Association of Universities for Research in Astronomy, 
Inc., under NASA contract NAS5-26555. Support for MAST for non-HST data is 
provided by the NASA Office of Space Science via grant NAG5-7584 and by other grants and contracts.


\begin{references}
\reference{a96} Abraham, R., Tanvir, N.R., Santiago, B.X., Ellis, R.S., Glazebrook,
K., \& van den Bergh, S. 1996, MNRAS, 279, L47
\reference{avb} Abraham, R., \& van den Bergh, S. 2001, Science, 293, 1273
\reference{a03} Abraham, R., van den Bergh, S., \& Nair, P. 2003, ApJ, 588, 218
\reference{sex} Bertin, E., \&  Arnouts, S. 1996,  A\&AS,. 117, 393
\reference{bo} Bouwens, R., Illingworth, G. D., Blakeslee, J.P., Broadhurst, T.J., \& Franx, M. 2004,
ApJ, 611, L1
\reference{c94} Calzetti, D., Kinney, A.L., \& Storchi-Bergman, T. 1994, ApJ, 429, 582
\reference{cas00}Casertano, S. et al. 2000 AJ, 120 2747
\reference{chap03}Chapman, S. C., Windhorst, R., Odewahn, S., Yan, H., \& Conselice, C. 
2003, ApJ, 599, 92
\reference{c04} Cimatti, A., et al. 2004, Nature, 430, 184
\reference{c03} Conselice, C. 2003, ApJS, 147, 1
\reference{cox05} Cox, T.J., Jonsson, P., Primack, J.R., \& Somerville, R. 2005, astro-ph:/0503201
\reference{d01} Daddi, E., Broadhurst, T., Zamorain, G., Cimatti, A., Rottering, H. \& Renzin, A. 2001,
A\&A, 376, 825
\reference{d03} Daddi, E. et al. 2004, 600, L127
\reference{d05} Dahlen, T. et al. 2005, ApJ, submitted.
\reference{d99} Dickinson, M. 1999,  in 
{\it After the Dark Ages: When Galaxies were Young (the Universe at 
2 $<$ z $<$5).} eds. S. Holt \& E. Smith, (College Park, Maryland: American 
Institute of Physics Press), p. 122
\reference{dp03}Dickinson, M., Papovich, C., Ferguson, H.C., \& Budavari, T. 2003, 
\reference{gd01} Giavalisco, M., \& Dickinson, M. 2001, ApJ 550, 177
\reference{g94} Giavalisco, M., Steidel, C., \& Szalay, A.S. 1994, ApJ, 425, 5
\reference{g96} Giavalisco, M., Steidel, C., \& Macchetto, F.D. 1996, ApJ, 470, 189
\reference{g02} Giavalisco, M. 2002, ARAA, 40, 579
\reference{g04} Giavalisco, M. et al. 2004, ApJL, 600, L1
\reference{g02} Goldader, J.D., Meurer, G., Heckman, T.M., Seibert, M., Calzetti, D., \& Steidel, C. 
2002, ApJ, 568,651
\reference{gkk} Gottl\"ober, S., Klypin, A., \& Kravstov, A.V.  2001, ApJ, 546, 223
\reference{f03} Ferguson, H.C., Dickinson, M., \& Papovich, C. 2002, ApJ, 569, L65
\reference{harry}Ferguson, H.C., et al. 2004, ApJL, 600, L107
\reference{hv} Hibbard, J.E.,  \& Vacca, W.D. 1997, AJ, 114, 1741
\reference{jon} Jonsson, P., Cox, T.J., Primack, J., \& Somerville, R. 2005, astro-ph:/0503135
\reference{k}Kannappan, S.J., Jansen, R., \& Barton, E. J. 2004, AJ, 127, 1371
\reference{lefev}Le F\`evre, O. et al. 2004, A\&A, 417, 839
\reference{lpm04} Lotz, J.M., Primack, J., \& Madau, P. 2004, AJ, 128, 163
\reference{l05}Lotz, J.M., et al., in prep
\reference{low}Lowenthal, J.D., et al. 1997, ApJ, 481, 673
\reference{mmw} Mao, H.J., Mo, S., \& White, S.D.M. 1998, MNRAS, 295, 319
\reference{m96} Madau, P., et al. 1996, MNRAS,  283, 1388
\reference{mh94}Mihos, C.J. \& Hernquist, L. 1994, ApJ, 425, L13
\reference{mh96}Mihos, C.J. \& Hernquist, L. 1996, ApJ, 464, 641
\reference{lexi}Moustakas, L. A., et al. 2004, ApJ, 600, L131
\reference{casey}Papovich, C. et al. 2004, ApJ, 600, L111
\reference{casey2}Papovich, C., Dickinson, M., Giavalisco, M., Conselice, C., \& Ferguson, H. 2005,
ApJ, in press, astro-ph/050188.
\reference{smn02} Steinmetz, M., \& Navarro, J. 2002, New Astronomy, 7, 155
\reference{steidel}Steidel, C., Giavalisco, M., Dickinson, M., \& Adelberger, K. 1996, AJ, 462, 17
\reference{spf} Somerville, R. S., Primack, J.R., \& Faber, S.M. 2001, MNRAS, 320, 504
\reference{toomre} Toomre, A. 1977, IAUS, 58, 347
\reference{vn} Vandeme, et al. 2004 in prep.
\reference{van}Vanzella, E. et al. 2005, A\&A, 434, 53
\reference{wirth}Wirth, G. D., et al. 2004, AJ, 127, 3121
\end{references}
\end{document}